\documentclass{article}     % onecolumn (ditto)
\usepackage{palatino}
\usepackage{mathpazo}
\usepackage{a4}
\usepackage{vmargin}
\setpapersize{A4}
\setmarginsrb{2cm}{1.5cm}{2cm}{1.5cm}{5mm}{5mm}{5mm}{5mm}

\usepackage{lineno}
\usepackage[section]{placeins} 

\usepackage{graphicx}

\usepackage{booktabs} % For formal tables

\usepackage{natbib}
\usepackage[caption=false]{subfig}
\usepackage{amsfonts,amsmath,amssymb,epsfig,tikz}
\usepackage{xspace}
\usepackage[strings]{underscore}
\usetikzlibrary{arrows,petri,topaths,decorations.pathreplacing,calc}
\usepackage{amsthm}
\theoremstyle{definition}

\newtheorem{example}{Example}
\usepackage{url}
\usepackage{tkz-berge}

\usepackage[automake,nonumberlist,acronym]{glossaries}
\makeglossaries
\newacronym{APTs}{APTs}{Advanced Persistent Threats}
\newacronym{FPOF}{FPOF}{Frequent Pattern Outlier Factor}

\newacronym{OD}{OD}{Outlier Degree}
\newacronym{OC3}{OC3}{One-Class Classification by Compression}
\newacronym{AVC}{AVC}{Attribute Value Coding} 
\newacronym{AVF}{AVF}{Attribute Value Frequency}
\newacronym{UI}{UI}{User Interface}
\newacronym{PE}{PE}{Process Event}
\newacronym{PX}{PX}{Process Exec}
\newacronym{PN}{PN}{Process Network}
\newacronym{PP}{PP}{Process Parents}
\newacronym{PA}{PA}{Process All}
\newacronym{nDCG}{nDCG}{Normalized Discounted Cumulative Gain}

\newacronym{ROC}{ROC}{Receiver Operator Characteristic}
\newacronym{AUC}{AUC}{Area Under Curve}

\newcommand{\AVF}{\textsc{AVF}\xspace}

\newcommand{\ProcessEvent}{\mathsf{PE}}
\newcommand{\ProcessExec}{\mathsf{PX}}
\newcommand{\ProcessParent}{\mathsf{PP}}
\newcommand{\ProcessNetflow}{\mathsf{PN}}
\newcommand{\ProcessAll}{\mathsf{PA}}

\renewcommand{\tfrac}[2]{{\textstyle \frac{#1}{#2}}}

\renewcommand{\cite}[1]{\citep{#1}}

\newcommand{\jrc}[1]{\textbf{\color{red}[James: #1]}}

\def\rot{\rotatebox}
\def\wbox{\parbox{4cm}}
\def\wboxh{\parbox{2cm}}
\newcommand{\ndcgPEI}{
\begin{tabular}{lccccc}
\toprule
& \textbf{FPOF} & \textbf{OD} & \textbf{OC3} &\textbf{CompreX} &\textbf{AVF} \\ \midrule
\textbf{Windows} &  0.20        &0.20        &  0.30             & \textbf{0.60}                                                              &\textbf{0.60}        \\ \midrule
\textbf{BSD}     &0.20        & 0.19     & 0.43              &                                                            \textbf{0.54} & 0.51         \\ \midrule
\textbf{Linux}   &  0.18        &0.18       &\textbf{0.38}   
                                                              & 0.30  &0.27       \\ \midrule
\textbf{Android} &  0.29       & 0.33        &  0.74 & 0.82           &   \textbf{0.84}    \\ \bottomrule
\end{tabular}
}

\newcommand{\ndcgPXI}{
\begin{tabular}{lccccc}
\toprule
                 & \textbf{FPOF} & \textbf{OD} & \textbf{OC3} &
                                                                \textbf{CompreX} & \textbf{AVF} \\ \midrule
\textbf{Windows} &  0.15        & 0.15       &\textbf{0.28}  &DNF             &    \textbf{0.28}      \\ \midrule
\textbf{BSD}     & 0.15       & 0.15     & \textbf{0.49}         &DNF        &  0.34       \\ \midrule
\textbf{Linux}   & 0.18         &0.18       & 0.30     &DNF             &\textbf{0.43}       \\ \midrule
\textbf{Android} &0.22         &  0.22       & \textbf{0.39} & DNF & \textbf{0.39}      \\ \bottomrule
\end{tabular}
}

\newcommand{\ndcgPPI}{
\begin{tabular}{lccccc}
\toprule
                 & \textbf{FPOF} & \textbf{OD} & \textbf{OC3} &
                                                                \textbf{CompreX} & \textbf{AVF} \\ \midrule
\textbf{Windows} &    0.10      & 0.10       & \textbf{0.21}     &DNF           &\textbf{0.21}          \\ \midrule
\textbf{BSD}     &  0.13      & 0.13     &  \textbf{0.43}     &DNF           & 0.30        \\ \midrule
\textbf{Linux}   &  0.17        &0.17       &\textbf{0.24}     &DNF              &0.20       \\ \midrule
\textbf{Android} & NA        &  NA      &  NA  &NA&NA      \\ \bottomrule
\end{tabular}
}

\newcommand{\ndcgPNI}{
\begin{tabular}{lccccc}
\toprule
                 & \textbf{FPOF} & \textbf{OD} & \textbf{OC3} &
                                                                \textbf{CompreX} & \textbf{AVF} \\ \midrule
\textbf{Windows} & 0.36         &0.36        & \textbf{0.71}    &DNF            &0.58     \\ \midrule
\textbf{BSD}     &  0.13      &0.14      &\textbf{0.32}     &DNF             &   0.26      \\ \midrule
\textbf{Linux}   &  0.23        & 0.23      &\textbf{0.48}  &DNF                 & 0.31      \\ \midrule
\textbf{Android} &      0.42   &0.36         &\textbf{0.67}  &DNF              &  0.47     \\ \bottomrule
\end{tabular}
}

\newcommand{\ndcgPAI}{
\begin{tabular}{lccccc}
\toprule
                 & \textbf{FPOF} & \textbf{OD} & \textbf{OC3} &
                                                                \textbf{CompreX} & \textbf{AVF} \\ \midrule
\textbf{Windows} &  DNF        &   DNF     &   DNF      &DNF        &  \textbf{0.52}        \\ \midrule
\textbf{BSD}     &  0.21      &0.19      & \textbf{0.65} &DNF  &  0.52       \\ \midrule
\textbf{Linux}   & DNF         &DNF     &DNF    &     \textbf{0.46}            &0.29       \\ \midrule
\textbf{Android} &   0.31      &   0.34      &  0.64  &DNF            &\textbf{0.83}       \\ \bottomrule
\end{tabular}
}

\newcommand{\ndcgPEII}{
\begin{tabular}{lccccc}
\toprule
                 & \textbf{FPOF} & \textbf{OD} & \textbf{OC3} &
                                                                \textbf{CompreX}  & \textbf{AVF} \\ \midrule
\textbf{Windows} &  DNF        &DNF        &  \textbf{0.23}   & \textbf{0.23}         &  0.21        \\ \midrule
\textbf{BSD}     &0.13       & 0.17     & \textbf{0.24}      & 0.21    & 0.19        \\ \midrule
\textbf{Linux}   &  0.22        &0.21       &0.38  & \textbf{0.46}               &0.29       \\ \midrule  
\textbf{Android} &  0.36       & 0.22        &  0.32 & \textbf{0.78}        &   0.30    \\ \bottomrule
\end{tabular}
}

\newcommand{\ndcgPXII}{
\begin{tabular}{lccccc}
\toprule
                 & \textbf{FPOF} & \textbf{OD} & \textbf{OC3} &
                                                                \textbf{CompreX} & \textbf{AVF} \\ \midrule
\textbf{Windows} & DNF        & DNF      &\textbf{0.24}  &DNF               &    \textbf{0.22}      \\ \midrule
\textbf{BSD}     & 0.18       & 0.17     & \textbf{0.51}      &DNF           &  0.17      \\ \midrule
\textbf{Linux}   & 0.20         &0.20       & \textbf{0.42}     &DNF             &\textbf{0.42}       \\ \midrule
\textbf{Android} &0.29       &  0.29       & \textbf{0.39}     &DNF          & \textbf{0.38}      \\ \bottomrule
\end{tabular}
}

\newcommand{\ndcgPPII}{
\begin{tabular}{lccccc}
\toprule
                 & \textbf{FPOF} & \textbf{OD} & \textbf{OC3} &
                                                                \textbf{CompreX} & \textbf{AVF} \\ \midrule
\textbf{Windows} &    DNF      & DNF       & \textbf{0.22}     &DNF           &\textbf{0.22}          \\ \midrule
\textbf{BSD}     &  0.10      & 0.09     &  \textbf{0.29}      &DNF          & 0.17        \\ \midrule
\textbf{Linux}   &  0.20        &0.20       &\textbf{0.42}     &DNF              &0.25       \\ \midrule
\textbf{Android} & 0.20       &   0.20     &             \textbf{0.39}&DNF   &0.25      \\ \bottomrule
\end{tabular}
}

\newcommand{\ndcgPNII}{
\begin{tabular}{lccccc}
\toprule
                 & \textbf{FPOF} & \textbf{OD} & \textbf{OC3}  &
                                                                \textbf{CompreX}& \textbf{AVF} \\ \midrule

\textbf{Windows} & DNF         &DNF        & DNF  &DNF    &DNF          \\ \midrule
\textbf{BSD}     & DNF  &\textbf{0.15}      &DNF   &DNF   &   DNF     \\ \midrule
\textbf{Linux}   &  DNF       & DNF      &DNF         & DNF  &DNF     \\ \midrule
\textbf{Android} &      0.37  &0.20         &\textbf{0.40}   &DNF             &  0.35    \\ \bottomrule
\end{tabular}
}

\newcommand{\ndcgPAII}{
\begin{tabular}{lccccc}
\toprule
                 & \textbf{FPOF} & \textbf{OD} & \textbf{OC3} &
                                                                \textbf{CompreX} & \textbf{AVF} \\ \midrule
\textbf{Windows} &  DNF        &   DNF     &   DNF     &DNF         & DNF \\ \midrule
\textbf{BSD}     &  0.21      &0.19      & \textbf{0.38}       &DNF          &  DNF \\ \midrule
\textbf{Linux}   & DNF         &DNF       &     \textbf{0.41}    &DNF          &DNF \\ \midrule
\textbf{Android} &   0.31      &   0.34      &  \textbf{0.82}     &DNF         &0.35 \\ \bottomrule
\end{tabular}
}

\begin{document}

\title{A baseline for unsupervised advanced persistent threat detection in system-level provenance}

\date{}

\author{Ghita Berrada \and 
        James Cheney \and
        Sidahmed Benabderrahmane \and
        William Maxwell \and
        Himan Mookherjee \and
        Alec Theriault \and
        Ryan Wright%etc.
}

\maketitle

\begin{abstract}
Advanced persistent threats (APT) are stealthy, sophisticated, and unpredictable cyberattacks that can steal intellectual property, damage critical infrastructure, or cause millions of dollars in damage.
%Why is it a problem?
Detecting APTs by monitoring system-level activity is difficult because manually inspecting the high volume of normal system activity is overwhelming for security analysts.
% Positive, startling statement
We evaluate the effectiveness of unsupervised batch and streaming
anomaly detection algorithms  over multiple gigabytes of provenance
traces recorded on four different operating systems to determine whether they can  detect realistic APT-like attacks reliably and efficiently. 
% Consequences
This report is the first detailed study of the effectiveness of
generic unsupervised anomaly
detection techniques in this setting.
\end{abstract}

\section{Introduction}
\label{sec:Intro}
 
For the past few years, damaging security/data breaches have frequently
made the
headlines~\cite{gootman2016opm,silver2014jpmorgan,lee2014german,Karchefsky2017}. These
breaches are all examples of ``advanced persistent threats'' (APTs).
\gls{APTs} are long-running, stealthy attacks designed to
penetrate specific target systems, carry out either pre-determined or
dynamically updated instructions from an adversary,
and persist (while avoiding detection) for as
long as required to accomplish the adversary's goals, such as data
theft~\cite{silver2014jpmorgan,gootman2016opm} or corruption of the
target organization's data and damaging of critical systems.

Security experts warn that APTs are now ``part and parcel of
doing business''~\cite{auty2015anatomy} and concede that it would be
unrealistic for all such attacks to be prevented and
blocked~\cite{smith2013life,maisey2014moving,auty2015anatomy}, partly
because even the best designed security systems are bound to have
flaws and partly because the targeted nature of the attacks means
that the adversaries will persistently try to gain access to the
target's system, adapting and changing their approaches if need be,
until they reach their goal or the cost of succeeding far outweighs
the benefits to be gained. As a result, the experts consider that, while adopting
state-of-the-art prevention techniques is a must, the focus should
shift to continuously monitoring the systems, detecting APTs in a
timely fashion and minimizing their damage.

Traditional security software and measures (e.g. anti-virus software,
system security policies) generally fail to detect APTs since APTs
tend to mimic normal business logic and rely on actions that respect
social norms (e.g. work schedule of targeted users) or system security policies. Moreover, the fact
that APTs are long-running campaigns that consist of multiple
steps further complicates their detection, in particular when relying
on event logs and audit trails that only provide partial information
on temporally and spatially localized events.

Provenance-tracking has been proposed as a basis for
security (e.g. provenance-based access
control~\cite{park2012provenance}). It has been suggested
that mining provenance data to analyze and identify causal
relationships among system activities could help identify security
threats and malicious actions, such as data exfiltration, that might
go undetected with policy-driven approaches and other classical
perimeter defence-based
methods~\cite{jewell2011host,zhang2012track,awad2016data,jenkinson2017applying}.

As appealing as the idea of monitoring provenance-like records to aid
security sounds, there are, however, numerous challenges to making it
a reality. Beyond the issues linked with recording the provenance
itself (e.g. level of provenance granularity, fault tolerance,
trustworthiness of the recorded trace~\cite{jenkinson2017applying}),
the recorded provenance traces are expected to be large in volume,
with anomalous system activity (if any) likely to constitute but a
very small fraction of the recorded traces. Analyzing provenance
traces to identify anomalous activity that would suggest an ongoing
APT attack is a typical ``needle in a haystack'' problem further
compounded by the variety of possible APT patterns and the lack of
available fully annotated data. Typical supervised learning techniques
cannot therefore be used to detect (rare) APT patterns\footnote{Training supervised learning models for the APT detection task would require having a corpus of provenance data with realistic APT attacks along with complete annotations indicating which parts of the provenance graphs are part of an attack. In an operational context, such annotations are not readily available and generating annotations for the provenance graphs a posteriori is prohibitively labor-intensive and time-consuming. Since we are developing/evaluating APT detection techniques to be used in an operational setting, we cannot assume the existence of a fully annotated corpus so this naturally precludes the use of supervised learning models. The high class imbalance inherent to this application also means supervised learning technques are not necessarily the best candidates for the detection task.}.  Furthermore,
unsupervised anomaly detection over streaming graphs is
challenging~\cite{graph-anomaly}. We know of only one paper on anomaly
detection over streaming provenance graph data~\cite{streamspot} but this
approach relies on an initial training stage over ``normal'' example
graphs, i.e. it is semisupervised.

In an operational security scenario, it is critical to be able to
provide actionable information quickly. Security analysts can usually
identify and forensically investigate suspicious behavior (such as
processes that have been subverted or created by an attacker) 
once it is brought to their attention.  However, in typical
system traces, each day of activity may lead to a gigabyte or more of
provenance trace information, corresponding to hundreds or thousands
of processes, almost all of which are benign. In this paper, we
consider the key subproblem of quickly identifying unusual process
activity that warrants manual inspection. Our approach summarizes
process activity using categorical or binary features such as the
kinds of events performed by a process, the process executable name
and parent executable name, and IP addresses and ports accessed.  We
focus on categorical data because attacks typically involve rare
combinations of such attributes.

This article evaluates the effectiveness of several algorithms for
unsupervised, categorical anomaly detection:
\begin{itemize}
\item FPOutlier (or \gls{FPOF})~\cite{fpoutlier}
\item Outlier Degree (or \acrshort{OD})~\cite{outlierdegree}
\item One-Class Classification by Compression (or \acrshort{OC3})~\cite{krimp-ad}
\item CompreX~\cite{comprex}
\item Attribute Value Frequency (or \acrshort{AVF})~\cite{avf,onepassavf}
\end{itemize}
All of these algorithms except for AVF are based on mining frequent
itemsets or association rules and using these results to assign
anomaly scores.  Moreover, these mining-based techniques are all batch
algorithms: in a first pass, the data is mined and analyzed (sometimes
taking a lengthy period) and in a second pass, the scores are assigned.
AVF is, instead, based on a simple analysis of the frequencies of the
attributes.  The original paper proposing AVF also only considered a
batch setting, but later work~\cite{onepassavf} showed how to modify
AVF to a one-pass, streaming algorithm.  We therefore refer to
\emph{batch} and \emph{streaming} AVF in this paper.

We apply our work to provenance traces containing example APT attacks
(on several different host operating systems) produced as part of the
DARPA Transparent Computing program, in which attacks constitute as
little as 0.01\% of the data.  We evaluated all of the above
algorithms in batch mode.  Our experiments show that on our dataset,
AVF has anomaly detection performance comparable or better than the
itemset mining-based techniques, typically finding at least some parts
of the attack within the top 1\% or even 0.1\%.

We also conducted experiments comparing batch and streaming AVF, using
a modified form of the one-pass algorithm of~\cite{onepassavf} that
allows blocks of different sizes, in order to study how detection
performance is affected by streaming.  Our experiments comparing batch
and streaming AVF with different block sizes show that there is little
degradation in anomaly detection performance.  Although our work (like
any anomaly-detection technique) does not guarantee to find all
attacks, our contribution demonstrates that unsupervised anomaly
detection can help find APT-style attacks that currently go unnoticed,
enabling analysts to focus their efforts where they are most needed.

This article does not propose new anomaly detection algorithms, and
does not evaluate all of the possible algorithms for unsupervised
anomaly detection on categorical data.  All of the algorithms
evaluated either have publicly-available implementations, or were easy
to re-implement.  It is possible that better results could be obtained
using other algorithms that we have not yet tried; nevertheless, our
results do establish a baseline against which new approaches (or
evaluation of other existing algorithms) can be measured. Such a
baseline is essential as a basis for assessing the effectiveness of
more sophisticated algorithms, and whether their complexity is
justified by increases in effectiveness.\\
The main contributions of this paper are:
\begin{itemize}
 \item Establishing baseline results for five categorical anomaly detection methods, i.e FPOF, OD, OC3, Comprex and AVF (in both batch and streaming modes for AVF) for the task of detecting APT-like activity in system provenance traces
 \item Thoroughly evaluating and comparing the effectiveness of these five anomaly detection methods for the studied task
 \item Showing that some methods, namely OC3 and AVF, already produce useful detection results in reasonable times despite their relative simplicity (``naive'' set of features requiring barely any domain knowledge or tweaking and/or very simple anomaly scoring strategy e.g. AVF) and that these results can, in some cases (e.g. AVF), very easily be replicated in a streaming setting
 \item Discussing appropriate metrics for the detection task and proposing a metric from information retrieval (normalized discounted gain) as a suitable metric
\end{itemize}
% \textless /end alternative proposal\textgreater

The structure of the rest of this paper is as follows.
Section~\ref{sec:overview} presents the overall system architecture
and outlines our approach. Section~\ref{sec:methods} reviews AVF and
our variant of streaming AVF.  % Section~\ref{sec:implementation}
% presents the details of our implementation.
Section~\ref{sec:evaluation} presents an experimental evaluation of
the effectiveness of the different approaches, establishing a baseline
for unsupervised anomaly detection on this data.
Section~\ref{sec:relatedwork} summarizes related work on APTs and
anomaly detection.  Section~\ref{sec:conclusion} concludes and
suggests directions for future work.

A short glossary of acronyms used in the paper is included as an appendix.

% Formal Concept Analysis
% (FCA) provides a principled way to explore provenance graph data,
% "summarize" it and try and discover "interesting" patterns and
% association rules in the data and uncover hidden relationships in the
% data. FCA would in particular allow us to learn normal behaviour
% through deriving association rules that occur with high confidence in
% benign data. Then, we would try and find violations of the derived
% rules under the assumption that entities that violate the rules are
% potentially suspicious and might be part of an APT attack.

% After briefly describing related work (Section \ref{sec:relatedWork})
% and explaining the basic principles of FCA and association rule mining
% (Section \ref{sec:introFCA}), we demonstrate how to use FCA and
% association rule mining to mine provenance graph data and uncover
% potentially suspicious patterns/entities in the data that could
% indicate an ongoing APT attack (Sections \ref{sec:anomaly} and
% \ref{sec:experiments}).

% \textbf{contrast with other techniques for AD over itemset data? Learn how to recognize familiar behavior from positive data that may have some infrequent negative data in it but we don't know where; able to explain why something is anomalous.}

\section{Overview}
\label{sec:overview}
 
\subsection{Provenance trace analysis}
\begin{figure}[tb]
%\begin{minipage}[b]{.5\textwidth}
  \begin{center}
   \resizebox{0.5\textwidth}{!}{
      \includegraphics[scale=0.4]{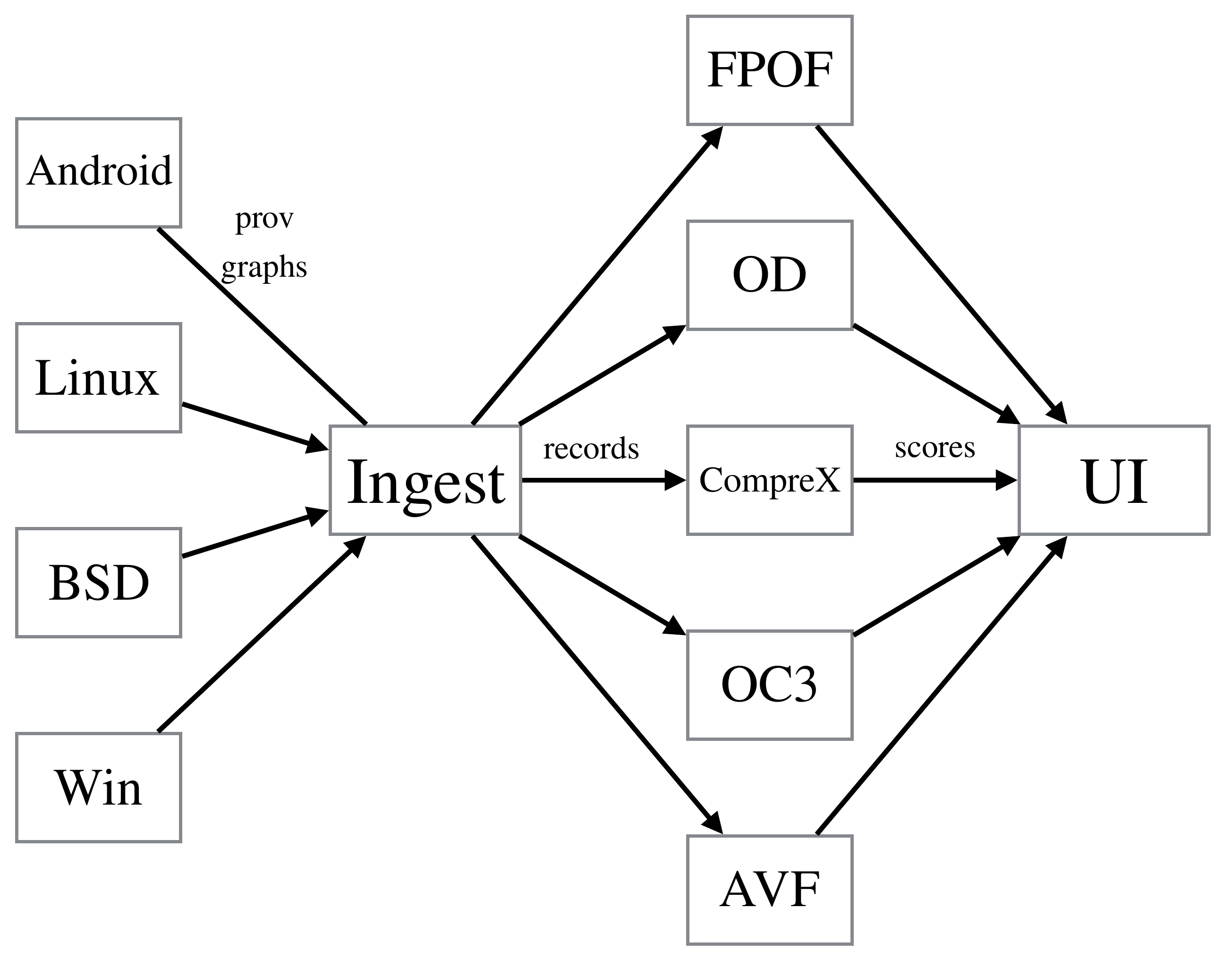}
    }
  \end{center}
  \caption{Architecture of our approach }\label{fig:arch}
\end{figure}

In this section, we situate our work as part of a realistic
provenance-based security scenario.  Figure~\ref{fig:arch} outlines
the architecture of our system, which is designed to interoperate with
several different \emph{(provenance)
  recorders}~\cite{gehani12middleware,jenkinson2017applying}, each
running on a different operating system and generating different
styles of provenance graphs recording system activity (albeit in a
common format). In this paper, we consider four sources, running on
Android, Linux, BSD and Windows operating systems.

%\vfill

Our system receives the provenance graph data from each recording
system, as a stream of JSON records in a binary format, and
\emph{ingests} the data into a graph database, Neo4J.  In addition,
ingestion performs some additional \emph{data integration} and
\emph{deduplication} steps to deal with some idiosyncrasies among the
sources.  The different systems use the shared data model in different
ways, for example storing information in different places, at
different levels of granularity, or just not populating some fields.
We remove some information that is not consistently recorded and
reorganize other information so that typical queries can be written
portably across data sources.  Deduplication is important because the
recorders add their own unique identifiers for operating system
processes and other objects.  This is necessary to avoid ambiguity
given that operating system-issued process identifiers or filenames
are not unique over long periods of time (i.e. days).  However, some
recording systems create multiple records referring to the same
process (or other object) with \emph{different} unique identifiers.
The ingester attempts to detect and merge these duplicates, using
heuristics such as ``two processes with the same process ID and
started at the same time are identical''.

Once the graph data has been ingested, we extract Boolean-valued
datasets called \emph{contexts} from the graph (an example of context is provided in Table~\ref{tab:ex_cxt}).  Each context
represents an aspect of process behavior as a Boolean-valued vector.
As a simple example, we could use attributes corresponding to event
types (\texttt{read}, \texttt{write}, etc.) with value `1' meaning
that the process performed at least one event of that type and `0'
otherwise; the exact number of such events is ignored.  We discuss
additional contexts later in this section.  Contexts can be extracted
using queries over the fully-ingested data, for forensic analysis, or
by incrementally maintaining appropriate data structures and
periodically emitting new records.  Each context can then be run
through the anomaly detection algorithms described in
Section~\ref{sec:methods}, yielding a score for each process.

\begin{figure}[tb]
  \begin{center}
    \resizebox{0.8\textwidth}{!}{
      \includegraphics{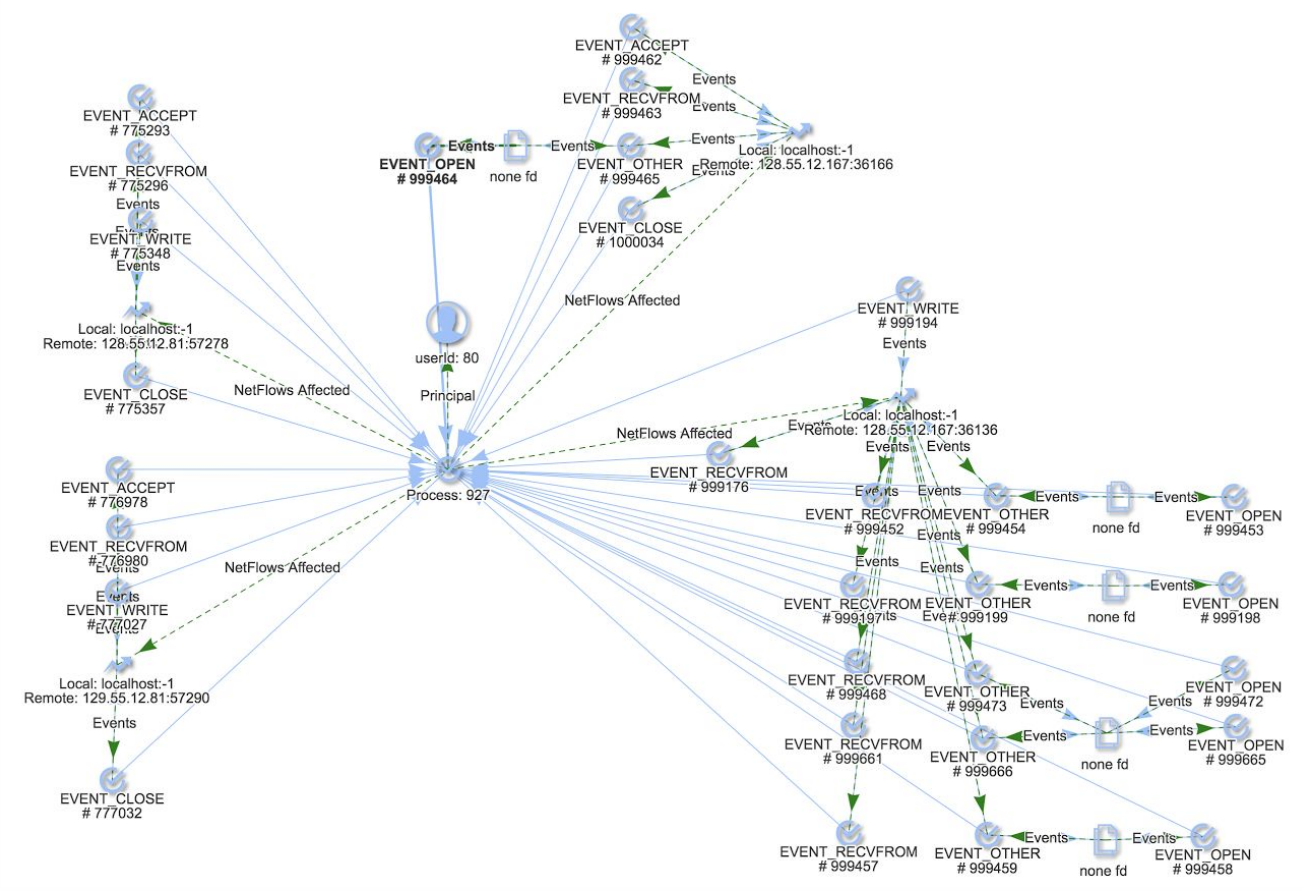}
   }
  \end{center}
  \caption{Example of attack provenance graph }\label{fig:prov}
\end{figure}

These scores are provided to the user interface (\gls{UI}) frontend, which
allows analysts to explore the graph using queries, or search for
anomalies based on the scores.  Figure~\ref{fig:prov} shows a typical
provenance graph created using the UI graph visualization system, as a
result of a successful attack detection.
This illustration highlights that even fairly simple
activities can yield complex graphs involving multiple
read/write or network access events.

Our system has participated in several DARPA exercises in concert with
the recording systems, in which realistic background activity was
simulated on each system, and realistic APT-style attacks were
performed, yielding several gigabytes of raw trace data, corresponding
to tens of millions of nodes and edges.  We have
manually annotated the data to indicate the processes constituting the
attacks for each of these scenarios.  Typically, the number of
processes involved in an attack is very small: for example, in the
largest dataset, there are over 282,000 processes (representing seven
days of activity), and only 46 of them (i.e. around 0.016\%) are
involved in the attack.  Even if we optimistically assume an analyst
can recognize an attack process in just 10 seconds, screening 200,000
processes would take over 23 days.  Thus, although attacks are often
easy to recognize once brought to the attention of an analyst, the
sheer volume of background activity makes it imperative to find ways
to automatically direct attention to suspicious activity.
\subsection{Contexts}
\label{subsec:contexts}
We now give the details of the contexts that form the starting point
for our proposed algorithms. In our approach, the  context definitions
are the \textit{only} places where domain knowledge about the  data is
used.  We consider the following contexts:

\begin{itemize}
\item ProcessEvent (\gls{PE}): The integrated traces use
  \emph{event types} such as \texttt{open}, \texttt{close},
  \texttt{exit}, etc.\ to describe process activity in a
  OS-independent way.  A process $p$ has attribute $ty$ if $p$ ever
  performs an event of type $ty$ (disregarding the exact number of events).
\item ProcessExec (\gls{PX}): The attributes are executable names $nm$,
  for example \texttt{ls} or \texttt{sudo}.  A process
  $p$ has attribute $nm$ if $p$ is an instance of executable $nm$.
\item ProcessParent (\gls{PP}): The attributes are again
  executable names $nm$.  A process $p$ has attribute $nm$ if $p$ is a
  child process of an executable named $nm$.
\item ProcessNetflow (\gls{PN}): The attributes are IP
  addresses $ip$ and port numbers $pn$.  A process $p$ has attributes
  $ip$ and $pn$ if it ever communicates with IP
  address $ip$ at port $pn$.
\item ProcessAll (\gls{PA}): the combination of all of the above
  contexts, with attributes renamed to avoid any ambiguity (for example
  between $\ProcessExec$ and $\ProcessParent$).
\end{itemize}
These contexts may seem rather simplistic.  For example, it seems
intuitive to also consider files accessed by processes as attributes.
Also, it would make sense to consider more complex attributes that
look for patterns that are known to be suspicious, such as
downloading a file, executing it, and then deleting it.  However, our
goal is to minimize the amount of fine-tuning needed to obtain useful
results.  There is also a trade-off between granularity of attributes
and performance: the more attributes we 
track, the more work needs to be done at each step.  Nevertheless, it
would be worthwhile, in subsequent work, to consider richer contexts or well-chosen
attributes that encode domain knowledge about what activities are
suspicious.
It might also be interesting to consider features that extend existing contexts, for example:
\begin{itemize}
 \item the number of times each type of event is performed or the
   frequency of each type of event performed (as opposed to just
   whether particular types of event are performed as in
   $\ProcessEvent$)
 \item Netflow properties not taken into account in $\ProcessNetflow$
   such as total number of bytes transferred
\end{itemize}
Such features would require discretization if they are to be used with the categorical anomaly detection methods explored in this paper. Otherwise, they would have to be used with numerical anomaly detection methods yet to be explored, with the results of such methods then fused with those obtained from categorical anomaly detection methods. This is beyond the scope of the current paper and will be explored in future work.

% Each of these contexts can be thought of as a relational table with
% the process node ID as the key and a number of Boolean fields
% corresponding to the attributes.  % We can therefore combine contexts by
% joining them on the process unique ID field.  For example, we write
% $P_{\mathsf{EX}}$ for $\ProcessEvent \Join \ProcessExec$; we also write
% $P_{\mathsf{ALL}}$ for the join of all four contexts.
% We implicitly rename attribute names
% when joining two contexts to avoid confusion between them; for example
% both $\ProcessExec$ and $\ProcessParent$ might have an attribute
% \texttt{bash}, so before taking a join we need to rename this
% attribute to, say, \texttt{exec(bash)} and \texttt{parent(bash)}
% respectively.

Each of these contexts can also be extracted from the data
incrementally, as the data is ingested.  For each process encountered,
we construct an attribute vector with value 1 for each attribute the
process has (in a given context) and 0 otherwise.  The resulting
sequence of vectors constitutes a \emph{dataset}
$D = x^{(1)},\ldots,x^{(n)}$ which we use as the starting point for
the algorithms in the next section.
\begin{table*}[htp]
\caption{Example of context: process identifiers vs type of system events, i.e  ProcessEvent (PE) context (extracted from Android provenance graph)}
\label{tab:ex_cxt}
\resizebox{\textwidth}{!}{
\begin{tabular}{cccccccccccccccccccccc}
\toprule
\rot{90}{\wbox{\small{Object\_ID}}}&\rot{90}{\wbox{\small{EVENT\_CLONE}}}&\rot{90}{\wbox{\small{EVENT\_CHECK\newline\_FILE\_ATTRIBUTES}}}&\rot{90}{\wbox{\small{EVENT\_OTHER}}}&\rot{90}{\wbox{\small{EVENT\_MPROTECT}}}&\rot{90}{\wbox{\small{EVENT\_CLOSE}}}&\rot{90}{\wbox{\small{EVENT\_CREATE\newline\_OBJECT}}}&\rot{90}{\wbox{\small{EVENT\_LSEEK}}}&\rot{90}{\wbox{\small{EVENT\_UNLINK}}}&\rot{90}{\wbox{\small{EVENT\_WAIT}}}&\rot{90}{\wbox{\small{EVENT\_MODIFY\newline\_PROCESS}}}&\rot{90}{\wbox{\small{EVENT\_RECVFROM}}}&\rot{90}{\wbox{\small{EVENT\_MODIFY\newline\_FILE\_ATTRIBUTES}}}&\rot{90}{\wbox{\small{EVENT\_WRITE}}}&\rot{90}{\wbox{\small{EVENT\_BIND}}}&\rot{90}{\wbox{\small{EVENT\_READ}}}&\rot{90}{\wbox{\small{EVENT\_RENAME}}}&\rot{90}{\wbox{\small{EVENT\_OPEN}}}&\rot{90}{\wbox{\small{EVENT\_LOADLIBRARY}}}&\rot{90}{\wbox{\small{EVENT\_CONNECT}}}&\rot{90}{\wbox{\small{EVENT\_SENDTO}}}&\rot{90}{\wbox{\small{EVENT\_SENDMSG}}}\\
\midrule
&&&&&&&&&&&&&&&&&&&&&\\
\wboxh{285d5fed-06dc-32ae\newline-a04a-13cc9426616b}&0&1&1&0&1&1&0&1&0&0&0&1&1&0&1&1&1&1&0&0&0\\
&&&&&&&&&&&&&&&&&&&&&\\
\wboxh{1e3548c0-b030-3591\newline-97ac-71b67bbcb305}&0&1&1&0&1&1&0&0&0&0&0&1&1&0&1&0&1&0&0&0&0\\
&&&&&&&&&&&&&&&&&&&&&\\
\wboxh{b4f1724e-0ba1-316b\newline-973f-69e5d5e3490c}&0&1&1&0&1&1&0&1&0&0&0&1&1&0&1&1&1&0&0&0&0\\
&&&&&&&&&&&&&&&&&&&&&\\
\wboxh{e2a4e818-3ce2-3626\newline-8e22-134b542d1d77}&0&0&1&0&0&0&0&0&0&0&0&0&1&0&1&0&0&0&0&0&0\\
......&&&&&&&&&&&&&&&&&&&&&\\
&&&&&&&&&&&&&&&&&&&&&\\
\bottomrule
 \end{tabular} }
 \end{table*}

\section{Algorithms}
\label{sec:methods}

We consider datasets $D$ to be sequences of $m$-dimensional Boolean vectors, where
there are $n>0$ vectors and $m>0$ attribute values.  Likewise, we
consider data sources to be streams of $m$-dimensional vectors.  In
either case, we consider a typical record $x^{(i)}$ at position $i$
and write $x^{(i)}_j$ for the value of attribute $j$ in $x^{(i)}$.
We assume for simplicity that all attributes are Boolean-valued. It is
not difficult to generalize to finite sets of attribute values.  We
also assume that the number of possible attributes $m$ is fixed.
% Without loss of generality, we assume that the attributes are organized such that for
% attributes $j_1<j_2$, if attribute $j_1$ occurs first in
% record $x^{(i_1)}$ and $j_2$ occurs first in
% record $x^{(i_2)}$ then $i_1 \leq i_2$.  

We start by reviewing the various batch-only approaches then describe both Attribute Value Frequency algorithm version, the 
original batch Attribute Value Frequency (AVF)
algorithm~\cite{avf} and its extension to a streaming setting.
We present the original algorithm in a batch
processing form, i.e. where we assume we have all of the data
before computing scores.  We show how to modify it to
obtain an online algorithm that gives a good approximation of the
results of the batch algorithm, and allows for a choice of different
window sizes.  This algorithm is a mild variation of the one-pass AVF
algorithm~\cite{onepassavf}.

\subsection{Batch-only anomaly detection techniques}
In this section, we briefly review the batch-only algorithms for anomaly
detection in the literature used in our evaluation.  These
descriptions are not exhaustive; the respective research papers should
be consulted for full details.

\subsubsection{FPOutlier (FPOF)}
The FPOutlier algorithm~\cite{fpoutlier} starts by mining frequent itemsets according to a
support parameter $minsupp$ (the algorithm only mines and considers itemsets that occur in a fraction of data transactions higher or equal to $minsupp$).  Then each object is assigned a score
corresponding roughly to the number of frequent itemsets it contains.
Thus, larger scores correspond to more occurrences of frequent
itemsets, meaning that anomalous objects should have low scores.  This
approach seems well-suited to detect anomalies corresponding to
expected, but missing, activity.  However, objects that have unusual
activity but also display a large number of common patterns may have
high scores and not be considered anomalous.  In addition, the fact
that this approach has a tunable parameter is problematic in an
unsupervised setting, since it means that we need to guess an
appropriate value for this parameter in advance.  We reimplemented
FPOutlier using standard itemset mining libraries.

\subsubsection{Outlier Degree (OD)}
The Outlier Degree algorithm~\cite{outlierdegree} also starts by mining frequent itemsets as well as
high-confidence rules, so there are two parameters, $minsupp$ governing
the minimum support of the itemsets and $minconf$ governing the minimum
confidence of the rules.  Then each object is scored by applying the
high-confidence rules to it, and assigning a score corresponding
roughly to the difference between the object's actual behavior and
expected behavior (according to the rules).  For example, if $X \to Y$
is a high-confidence rule and object $O$ displays behavior $X$ but not
$Y$, this will contribute to the score.  High scores correspond to
larger differences between actual and expected behavior, so are more
anomalous.  Like FPOutlier, this approach seems more likely to
consider missing, but expected, behaviors to be anomalous, and could
miss anomalies that consist of rare behaviors that do not occur
frequently enough to participate in rules.  Also, the presence of two
tunable parameters is even more problematic from the point of view of
unsupervised anomaly detection.  We reimplemented OD using standard
itemset and rule mining libraries.

\subsubsection{One-Class Classification by Compression (OC3)}

OC3~\cite{krimp-ad} is based on a compression technique for
identifying ``interesting'' itemsets, implemented using the Krimp
algorithm~\cite{krimp}.  Essentially, the idea is to first
mine frequent itemsets from the data, and then identify a subset of
the itemsets that help to compress the data well.  Then, each object
is assigned an anomaly score corresponding to its estimated compressed
size.  If the compression algorithm has done a good job, then objects
exhibiting commonly occurring patterns will compress well, and
anomalies will not. So high compression sizes (i.e high scores) point to anomalies.  OC3 can take a $minsupp$ support parameter, but
parameter tuning is typically not necessary because the compression
algorithm will filter out any non-useful itemsets; therefore we used
the smallest possible $minsupp$ setting in our experiments.  The
implementation of Krimp is available and we modified it slightly to
perform OC3-style anomaly scoring.

\subsubsection{CompreX}

CompreX~\cite{comprex} is perhaps the most sophisticated
approach studied to date.  It is based on compression, like OC3, but
uses a different compression strategy.  CompreX searches for a
partition of the attributes such that each set of attributes in the
partition has high mutual information.  Since there are
exponentially many partitions to consider, CompreX starts with the
finest partition (all attributes are in their own class) and greedily
searches for pairs of classes to merge.  Each resulting partition is
then compressed separately, to obtain an anomaly score for each record
based on its compressed size, as in OC3.  CompreX has no tuning
parameters and was shown experimentally to be competitive or superior
in anomaly detection performance to Krimp/OC3 on several datasets.
However, CompreX's default search strategy is quadratic in the number
of attributes; therefore, it was not usable on contexts with over
20-30 attributes.

\subsection{Attribute Value Frequency (AVF)}
In this section, we describe the original batch Attribute Value
Frequency (AVF) algorithm~\cite{avf} and then its modification to suit a
streaming setting~\cite{onepassavf}.  Unlike the algorithms mentioned earlier, AVF is
rather simple and does not require additional background material
to describe, both in the batch and streaming settings.  Since we
implemented both variants of AVF from scratch in a unified way, rather
than reusing existing libraries or implementations as for the other
approaches, we will spell out the details.

Attribute Value Frequency (AVF)~\cite{avf} is a non-parametric outlier
detection technique appropriate for categorical data and was shown to
be fast, scalable and accurate on a variety of standard data sets. The
algorithm relies on the intuition that outliers in a dataset have
values of attributes which occur infrequently. That the
attribute values in a data point are infrequent can be determined simply
by computing the frequencies of the respective attribute values across
the data.  

Given a dataset $D$ of size $n$, we write $c_j$ for the number of
occurrences of attribute value $1$ for attribute $j$, i.e.
$c_j = |\{i \mid x^{(i)}_{j} = 1\}| = \sum_{i=1}^n x^{(i)}_{j}$.
Then, the AVF score of a data point $x$ is:
{
\[\AVF(x) = \frac{1}{m}\sum_{j=1}^{m} (x_j c_j + (1-x_j) (n-c_j))\]
}%
That is, when $x_j = 1$, the contribution to the score for attribute
$x_j$ is $c_j$, the number of occurrences of $j$-value of 1, and when
$x_j = 0$, the contribution is the number of occurrences of a
$j$-value of 0.  The initial multiplication by $1/m$ effectively
averages the counts, so $0 \leq AVF(x) \leq n$, but such scaling has
no effect on the relative ordering among scores in the batch setting.
Lower AVF scores indicate more unusual behavior.

\begin{example}[Running example]
  To illustrate AVF, we introduce a small running example
  with four processes $P_{17}, P_{42}, P_{1337}, P_{007}$ and three attributes
  $abc.com$, $xyz.com$ and $evil.com$, corresponding to network
  addresses accessed by the processes.  In this (extremely simplistic)
  example, $P_{17}$ and $P_{42}$ are innocuous activity and access
  both \texttt{abc.com} and \texttt{xyz.com}, while
  $P_{1337}$ is a naive attacker that only accesses \texttt{evil.com}
  and $P_{007}$ is a more sophisticated attacker that accesses all
  three in order to attempt to camouflage its behavior.  This behavior corresponds to the following dataset:
  {
\[\begin{array}{c|ccc}
id & \texttt{abc.com} & \texttt{xyz.com} & \texttt{evil.com}\\\hline
P_{17} & 1 & 1 & 0\\
P_{42} & 1 & 1 & 0\\
P_{1337} & 0 & 0 & 1\\
P_{007} & 1 & 1 & 1 
\end{array}\]
}

We calculate the frequencies of the
  three attributes as $c_{\texttt{abc.com}} = c_{\texttt{xyz.com}} =
  3$ and $c_{\texttt{evil.com}} = 2$.  Thus, the AVF scores
are:
  {  \begin{eqnarray*}
    \AVF(P_{17})  &=& \tfrac{1}{3}(3 + 3 +2) = \tfrac{8}{3}\\
    \AVF(P_{42})  &=& \tfrac{1}{3}(3+ 3 + 2) = \tfrac{8}{3}\\
    \AVF(P_{1337})  &=&  \tfrac{1}{3}(1+1+2) = \tfrac{4}{3}\\ 
    \AVF(P_{007})  &=& \tfrac{1}{3}(3+3+2) =\tfrac{8}{3}
  \end{eqnarray*}}%
The naive attacker's isolated access of \texttt{evil.com},
together with failure to mask its activity with common behavior,
results in a lower score, while the more
sophisticated attacker's score is the same as that of the first two processes.
\end{example}

\paragraph{Streaming AVF: Naive approach}
A simple, but unfortunately too naive, approach to streaming the AVF
algorithm is to maintain the attribute value counts incrementally as
data is processed, and use the current counts to score each new
transaction.  That is, if $c^{(i)}_j$ are the counts calculated for
$x^{(1)}\ldots x^{(i)}$, then to score a new record $x = x^{(i+1)}$ we
proceed as follows: {
\[\AVF_{naive}^{(i)}(x) = \frac{1}{m}\sum_{j=1}^{m} (x_j c^{(i)}_j + (1-x_j) (i-c^{(i)}_j))\]
}%
However, because the counts are monotonically increasing, this means
that the scoring will be heavily biased towards considering records
appearing early in the dataset to be anomalous.  For example:

\begin{example}
  Continuing our running example, we need to update the
  counts after each step.  Thus, the AVF scores
  are:
  {
  \begin{eqnarray*}
    \AVF(P_{17})  &=& \tfrac{1}{3}(0+0+1) = \tfrac{1}{3}\\
    \AVF(P_{42})  &=& \tfrac{1}{3}(1 + 1 + 2) = \tfrac{4}{3}\\
    \AVF(P_{1337})  &=&  \tfrac{1}{3}(1+1+0) = \tfrac{2}{3}\\ 
    \AVF(P_{007})  &=& \tfrac{1}{3}(2+2+1) = \tfrac{5}{3}
  \end{eqnarray*}
}
In this (admittedly extreme) example, the first process $P_{17}$ is judged
most anomalous, followed by $P_{1337}$, then $P_{42}$ and finally $P_{007}$.
\end{example}

\paragraph{Streaming AVF}
As observed by~\cite{onepassavf}, the problem is that the ``scale'' of the AVF scores is
not fixed in the streaming setting, since seeing an attribute whose
value has occurred only once means something very different for the
5th record in the dataset than for the 5000th record.

Instead, to compute AVF-like scores incrementally, we propose to
use the frequency counts to estimate probabilities for each attribute.
 We initially take $p^{(0)}_j = 0$ since the data is
typically sparse (having relatively few attribute values $x_j = 1$);
however, any other initial probability distribution could be used
based on domain knowledge.
%We will maintain the frequencies $c^{(n)}_j$ of
%observing each attribute $j$ to have value 1; we can easily do this
%incrementally by taking
% %
% {
% \[ c^{(0)}_j = 0 \qquad c^{(n+1)}_j = x^{(n+1)}_j + c^{(n)}_j\]
% }
% %
Next, for each new record $x^{(i+1)}$, we adjust
the probability $p_j^{(i+1)}$ of each attribute value $j$ being 1 after seeing
$x^{(i+1)}$ as follows:
{\small
\[p^{(i+1)}_j = \tfrac{n\times p^{(i)}_j+x_j}{i+1} \]
}%
% where $c^{(n)}_j$ is the number of observations of attribute $j$
% having value 1 in the first $n$ observations. We start with a
% uniform distribution $p^{(0)}_j=0$ for each attribute $j$.  
We then calculate the AVF score for
the $i+1$st record $x = x^{(i+1)}$ as follows:
{
\[\AVF^{(i+1)}(x) = \frac{1}{m}\sum_{j=1}^{m} (x_j p^{(i+1)}_j + (1-x_j) (1-p^{(i+1)}_j))\]
}%
%
% Note that each score is calculated just once; we do not incrementally
% re-calculate previous scores when new data is seen.  Instead we score
% each new record based on the information available at the time. Thus,
% the online version of AVF will generally produce different results
% than the batch version.
Note that, in the batch setting, dividing the counts by $n$ and summing
probabilities instead of counts would not affect the final results,
because all the counts are divided by the same $n$.  However, for the
streaming setting, we update the attribute value probabilities after
each step, so the results of AVF scoring will be different in the
streaming setting.  

% We want to emphasize that, while the (naive) offline version of the AVF
% algorithm described above is prior work (due to \citet{avf}), this
% prior work did \emph{not} discuss how to apply it to streaming data.
% In fact, directly using the the original version of AVF would not
% work: it actually averages the \emph{raw counts} of the attributes, not
% their \emph{frequencies} (probability estimates).  In the batch case,
% this makes no difference because the scores obtained by summing counts
% are the same as those obtained by summing frequencies, up to a
% constant scaling factor ($1/n$).  However, in the online setting,
% summing the counts makes little sense, because it means that the
% resulting scores will (generally) increase over time as new data is
% seen.  We experimented with this and found that it is nearly useless
% for anomaly detection.  Therefore, we take the alternative approach of
% averaging probabilities, which (as we will see) does produce
% reasonable results.

\begin{example}
  Continuing our running example, we now update the
  probabilities after each step.  Thus, the AVF scores
  are:
{
  \begin{eqnarray*}
    \AVF(P_{17})  &=& \tfrac{1}{3}(0+0+1) = \tfrac{1}{3}\\
    \AVF(P_{42})  &=& \tfrac{1}{3}(\tfrac{1}{2} + \tfrac{1}{2} + 1) = \tfrac{2}{3}\\
    \AVF(P_{1337})  &=&  \tfrac{1}{3}(\tfrac{1}{3} + \tfrac{1}{3} + 0) = \tfrac{2}{9}\\
    \AVF(P_{007})  &=& \tfrac{1}{3}(\tfrac{1}{2} + \tfrac{1}{2} +
                       \tfrac{1}{4}) = \tfrac{5}{12}
  \end{eqnarray*}
}%
  % {\[
%   \begin{array}{rclclcrclcl}
%     \AVF(P_{17})  &=& \tfrac{1}{3}(0+0+1) &=& \tfrac{1}{3}&\qquad &
%     \AVF(P_{1337})  &=&  \tfrac{1}{3}(\tfrac{1}{3} + \tfrac{1}{3} + 0) &=& \tfrac{2}{9}\\
%     \AVF(P_{42})  &=& \tfrac{1}{3}(\tfrac{1}{2} + \tfrac{1}{2} + 1) &=& \tfrac{2}{3}&&
%     \AVF(P_{007})  &=& \tfrac{1}{3}(\tfrac{1}{2} + \tfrac{1}{2} +
%                        \tfrac{1}{4}) &=& \tfrac{5}{12}
%   \end{array}
% \]}%
The naive attacker's behavior results in a lower (more anomalous)
score than the first process $P_{17}$.
\end{example}

\subsubsection{Analysis}

As outlined already, the batch AVF approach is implementable as two scans
over the data, and the online AVF approach can be implemented in a single,
linear scan, where scoring each new record and updating the
frequencies takes $O(m)$ time and space.  Both algorithms just need
to maintain the number of records $n$ and the $m$ counts or
probabilities.  Thus, the overall time complexity of each algorithm is
$O(nm)$ and the space required is $O(m)$.  In our experiments, the
number of attributes $m$ ranges from around 20 to over 14,000.  Our
approach may not scale well if the attributes are fine-grained and $m$
is much larger than $n$.

Another concern the reader might have is regarding arithmetic
precision and overflow.  If fixed-size (say, 32-bit) integers are
used, then whenever we are in danger of overflowing we can rescale by
dividing all of the counts by 2; this is exactly what is done in
arithmetic coding~\cite{witten87cacm}.  Our implementation uses
arbitrary-precision arithmetic.

\section{Experimental evaluation}
\label{sec:evaluation}

\subsection{Experimental setup}

The experiments were run on a desktop with an Intel Core i7-6700 CPU
(3.4 GHz), 16 GB RAM, running Ubuntu 16.04.  The raw provenance trace
data was ingested on a variety of machines and the contexts
used in the experiments were extracted and stored as CSV
files\footnote{available at http://www.gitlab.com/adaptdata}.  We do
not report the experimental setup for the ingestion stage here in
detail; however, it is easily able to keep up with the data in
real-time (that is, ingestion of data representing 7 days of system
activity takes much less than 7 days).  Our experiments focus on evaluating
the detection effectiveness and runtime cost of the anomaly detection algorithms
on the given context data.

\subsection{Datasets}
In our experiments, we use two data collections described in Table~\ref{datatable} and representing two attack scenarios, each consisting of several days'
worth of activity in a DARPA evaluation of provenance-tracking systems, running
on Windows, BSD, Linux and Android respectively.
These data collections result from two exercises for evaluating provenance recorders and anomaly detection techniques. 
The first data collection/scenario (a) consists of roughly 5 days of processes and netflows activities,  
whereas the second data collection/scenario (b) corresponds to around 8 days of data generated in similar conditions to the previous scenario. 
The provenance graphs have been recorded on four different tracking systems, running on Windows, BSD, Linux and Android respectively, 
each of which was subject to (part of) an APT-style campaign. The main
differences between scenarios 1 and 2 concern the background activity workload, 
the quality and the robustness of the attacks, and the size of the
provenance graphs.

Table~\ref{datatable} records, for each triplet context (rows 3 to 7)/OS/scenario (OS and scenarios are columns), the number of transactions $n$ (top value per context row) and the number of attributes
$m$ (bottom value per context row).  
The number of processes encountered in each system varies
significantly: in particular, the Linux dataset records from 3--10
times as many distinct processes compared to the Windows or BSD
datasets and up to 2400 times as many processes compared to Android.
Some contexts are empty, e.g.  $\ProcessParent$ for Android in
Scenario 1,
where information about parent process relationships was unavailable.  In
general, among the base contexts, the $\ProcessEvent$ context usually
has the largest number of processes, followed by $\ProcessExec$ and
$\ProcessParent$, while $\ProcessNetflow$ or $\ProcessExec$ have the
largest number of attributes, followed by $\ProcessParent$.  The number of attacks per OS/scenario is extremely low and ranges from 8 (Windows both scenarios) to 46 (Linux scenario 2).
%There are 9 attack processes in the Android data (8.8\%), 8 in the Windows data (0.04\%), 13 in the BSD data (0.02\%) and 25 in the Linux data (0.01\%).
Note that the size of the original dataset does not
directly correlate with the number of processes or attributes.  For
example, in scenario 1, the Android dataset is the largest but has the fewest
processes and attributes, because the provenance recorder for Android
records a great deal of low-level app activity and dynamic information
flow tracking, which we do not analyze.
The last row represents the percentage of attacks observed in each OS/context. For example, there are 8 attack processes in the Windows data (0.04\%) in the first scenario, and 8 (0.07\%) in the second one. The percentage of attacks per OS/scenario goes as low as 0.004\% (BSD scenario 2) and as high as 8.8\% (Android scenario 1).

\begin{table*}
\centering
	\scriptsize

\caption{Description of the datasets used during the experiments. In each context row (rows 3 to 7), the element at the top shows the number of rows (processes) and the element at the bottom the number of columns (attributes).}

 \label{datatable}
\resizebox{\linewidth}{!}{
\begin{tabular}{lllllllll}
\toprule&\multicolumn{2}{c}{Windows}&\multicolumn{2}{c}{BSD}&\multicolumn{2}{c}{Linux}&\multicolumn{2}{c}{Android}\\
\cmidrule{2-9}
&\multicolumn{2}{l}{Scenario}&\multicolumn{2}{l}{Scenario}&\multicolumn{2}{l}{Scenario}&\multicolumn{2}{l}{Scenario}\\
&1&2&1&2&1&2&1&2\\
\midrule
Size&743&9.53&288&1.27&2858&25.9&2688&10.9\\
&MB&GB&MB&GB&MB&GB&MB&GB\\
\midrule
ProcessEvent&17569&11151&76903&224624&247160&282087&102&12106\\
(PE)&22&30&29&31&24&25&21&27\\
\midrule
ProcessExec&17552&11077&76698&224246&186726&271088&102&12106\\
(PX)&215&388&107&135&154&140&42&44\\
\midrule
ProcessParent&14007&10922&76455&223780&173211&263730&0&24\\
(PP)&77&84&24&37&40&45&0&11\\
\midrule
ProcessNetflow&92&329&31&42888&3125&6589&8&4550\\
(PN)&13963&125&136&62&81&6225&17&213\\
\midrule
ProcessAll&17569&11151&76903&224624&247160&282104&102&12106\\  
(PA)&14431&606&296&265&299&6435&80&295\\
\midrule
$nb\_attacks$&8&8&13&11&25&46&9&13\\
\midrule
$\%\frac{nb\_attacks}{nb\_processes}$ &0.04&0.07&0.02&0.004&0.01&0.01&8.8&0.10\\
\bottomrule
\end{tabular}
}
\end{table*}

\subsection{Evaluation metrics}

The anomaly detection methods that we evaluate output a ranking of processes according to
their degree of suspiciousness/anomaly scores. These methods do not
explicitly classify or label entities as anomalous or normal.
Moreover, the data is unbalanced, with between 0.004\% and 8.8\% of
the data belonging to attacks. A high accuracy could be obtained by
simpliy classifying all processes as non-attacks, so accuracy would be
a poor indicator of model quality: this is the accuracy
paradox~\cite{thomas2008}.  That being the case, it would not be
appropriate to use metrics usually employed to evaluate classification
methods.
% On top of that,
% the data is unbalanced (depending on the dataset,
% at most 8.8\% and as low as 0.004\% of the data points belong to the attack class i.e. are true positives), constraining our choice of metrics.
% Accuracy, in particular, would not be an appropriate metric as, given the extremely unbalanced nature of our data, a very high accuracy could be achieved---for an arbitrarily fixed threshold of
% processes---simply by classifying all samples as not being part of an attack. This would clearly be unacceptable.

\subsubsection{Normalized discounted cumulative gain}
To evaluate the anomaly detection algorithms described earlier, we propose using a metric called the normalized discounted cumulative gain metric (or nDCG for short). It is a metric often used in information retrieval to assess the quality of a ranking.

Given a typical document search application,
\citet{jarvelin2002cumulated} argued that, from a user's perspective,
relevant documents are more valuable to a user than marginally
relevant documents and a relevant document ranked high in the returned
list of results is more valuable than an equally relevant document
ranked lower in the list. A user may be reasonably assumed to scan the
list of returned results from the beginning before interrupting the
scan at some point correlated with time availability, effort required
as well as the cumulated information from documents already seen.  So
it is safe to assume that relevant documents located further down the
list of returned results are unlikely to be seen by the user as they
would require more time and effort and become less valuable. Taking
these facts into account, \citet{jarvelin2002cumulated} introduced the
nDCG measure.

We similarly argue that, in our application, processes that are part of an attack but are ranked very low by an anomaly detection
technique are virtually useless to an analyst since his/her monitoring burden would increase substantially with the amount of processes to be checked (not to talk about issues such as acquired loss of trust in
the automated monitoring system and discarding of its alerts as well as the increased potential for misses and errors with the increase of data to monitor). Because of this, we believe nDCG to be an appropriate metric for our application.

To compute the \acrshort{nDCG}, we start by computing a score called discounted cumulative gain or DCG. The basis of DCG is that each document/entity in the ranking is assigned a relevance score and is penalized by a value
logarithmically proportional to its position/rank in the list of results. The DCG is therefore computed as follows:
\begin{align*}
DCG_N=\sum_{i=1}^N\frac{rel_i}{\log_2(i+1)}
\end{align*}
where $N$ is the number of entities/documents in the list, $rel_i$ the
relevance score of the $i$-th entity/document in the list.  

Since the length of result lists can vary and the DCG score does not
take that into account, it is common to normalize the DCG score by the ideal DCG
score (iDCG), which is simply the best achievable DCG score, i.e. the
score that would be achieved if all relevant entities were at the top
of the list (and in the case of different degrees of relevance, with
the highest values of relevance at the very top). Assuming we have $p$
relevant entities in the list, we have:

\begin{align*}
 iDCG_N&=\sum_{i=1}^N\frac{rel_i}{\log_2(i+1)} &
 nDCG_N&=\frac{DCG_N}{iDCG_N}
\end{align*}

In our case, we only consider entities to be either relevant
(processes that are part of an attack) or irrelevant (processes with
normal behavior) and assign a relevance score $rel_i$ of 1 to attack
processes and of 0 to benign processes, and the idealized score
results from ranking all $k$ attack processes at positions
$1,\ldots,k$.
% So the iDCG score simplifies to:
% \begin{footnotesize}
% \begin{align*}
% iDCG_N=\sum_{i=1}^N\frac{1}{\log_2(i+1)}
% \end{align*}
% \end{footnotesize}
The closer the nDCG score to 1, the better the ranking.
\subsubsection{Area under curve}
The \emph{receiver operator characteristic} curve (or \acrshort{ROC} curve) for a given ranking
of objects plots the fraction of true positives found against the
number of false positives found.  The
 area under \acrlong{ROC} curve (also called \gls{AUC}) is often used as a measure of anomaly
detection performance

In our case, the AUC would correspond to the proportion of processes with
normal behavior ranked lower than processes that are part of an
attack, computed as follows:
\begin{align*}
 \frac{1}{|A||\overline{A}|}|\{(\alpha,\beta):r(\alpha)<r(\beta), (\alpha,\beta) \in A \times\overline{A}\}|
\end{align*}
where $A$ is the set of elements with a relevant label (i.e. elements that are part of an attack),
$\overline{A}$ is the set of elements with an irrelevant label (i.e. elements that have a normal behavior),
$r(\alpha)$ (resp. $r(\beta)$) is the rank assigned to $\alpha$ (resp. $\beta$) by the method to be evaluated.
The best performance for a method under this metric (resp. the worst
performance) is achieved with AUC of one (resp. of zero).

However, in the presence of sparse anomalies in
large datasets, the AUC score's usefulness is somewhat limited.  The AUC can
either overestimate the effectiveness of an algorithm (e.g. if all
attacks are found at rank 900--1000 out of 200,000 then the AUC will
be over 0.995 but the results are still nearly useless), or
underestimate it (e.g. if half of the attacks are found in the top 10
and the other half at rank 100,000, then the maximum AUC is around 0.75
even though these results might be very
valuable). \citet{tapp2019} reported some experiments on the same
datasets including both 
AUC values and nDCG scores for the OC3 and AVF algorithms and found
that the scores are loosely correlated but
AUC scores are typically uniformly high values and much higher than nDCG scores.  AUC scores
usually fell in the relatively narrow range 0.75--0.99 (which would seem to indicate that all algorithms perform well in the attack detection task),
whereas nDCG scores range typically from 0.2--0.8 (suggesting more nuanced performances).  Based on this, AUC values wouldn't necessarily allow to properly discriminate between well performing and poorly performing algorithms. We will therefore
present only the nDCG scores for the batch algorithms, but present both nDCG
and AUC scores for the comparison of batch and streaming AVF in order
to understand whether either metric is affected by stream processing
or block size.

\subsection{Forensic anomaly detection}

In this section we consider the following empirical question:
\begin{itemize}
\item Q1: Can the five batch methods (FPOF, OD, OC3, CompreX, AVF)
  detect APT-style attacks effectively?
\end{itemize}

We first evaluate the effectiveness and performance of the batch
version of AVF compared with several other offline techniques, such as
FPOutlier (FPOF)~\cite{fpoutlier}, Outlier-degree (OD)~\cite{outlierdegree},
OC3~\cite{krimp-ad}, and CompreX~\cite{comprex}.

% FPOF~\cite{fpoutlier} is an itemset-mining approach that measures the outlierness of a transaction. If a transaction contains fewer frequent patterns, then the transaction will have small FPOF values and is likely to be an outlier. 
% OD~\cite{outlierdegree} considers outlier detection with both categorical and continuous variables. Transactions with infrequent values would indicate outliers. 
% OC3~\cite{krimp-ad} is another technique based on pattern based
% compression of the data~\cite{krimp}; since it uses the MDL principle
% by extracting the set of frequent itemsets that best compresses the
% input database and then assigning anomaly scores based on compressed size.

FPOF and OD were reimplemented in Python according to the descriptions
of the algorithms.  We reused publicly-available implementations of
OC3 and CompreX\footnote{http://eda.mmci.uni-saarland.de/prj/},
implemented in C++ and Matlab respectively.  The FPOF, OD and OC3
methods require setting some parameters, which is not the case for AVF
or CompreX.  For OC3, we used the lowest possible support parameter
and used closed itemset mining to reduce the total number of itemsets
considered in the mining stage.  For FPOF and OD, we considered a
range of support and confidence parameter
settings in the range 0.1--0.9 and 0.97 and report the best results obtained using any
parameter setting.

We report the results of all algorithms running on the contexts described in Section~\ref{subsec:contexts} in Table~\ref{tab:ndcgScenarioI} %\ref{tab:batch-evaluation-pe-1s}-\ref{tab:batch-evaluation-pa-1s}
for the first scenario, and Table \ref{tab:ndcgScenarioII} %\ref{tab:batch-evaluation-pe-2s}-\ref{tab:batch-evaluation-pa-2s} 
for the second one.  Some algorithms did not finish
within a reasonable time  (more than 3 hours) and when this is the
case we write $DNF$.  This happens most often with CompreX on contexts
where there
are large numbers of attributes, because CompreX
searches for a partition of the attributes into groups with high
mutual information, which seems to exhibit quadratic running time in
the number of attributes.

\begin{table}[tb] 
\caption{Evaluation of batch anomaly scoring in Scenario
  1 (nDCG scores). The higher the score (i.e the closer to 1) the better. The best score per OS (row) is highlighted in bold.}
\centering
\subfloat[ProcessEvent]{ \resizebox{0.7\linewidth}{!}{ \ndcgPEI \label{tab:batch-evaluation-pe-1s}}}
%%%%%%%%%%%%%%%%%%%%%%%%%%%%%%%%%%%%%%%%%%%%%%%%%%%%%%%%%%%%%%%%%%%%%%%%%%%%%%
  
\subfloat[ProcessExec]{ \resizebox{0.7\linewidth}{!}{\ndcgPXI \label{tab:batch-evaluation-px-1s} }}
\\
%%%
%%%%%%%%%%%%%%%%%%%%%%%%%%%%%%%%%%%%%%%%%%%%%%%%%%%%%%%%%%%%%%%%%%%%%%%%%%%%%%
%  
\subfloat[ProcessParent]{ \resizebox{0.7\linewidth}{!}{ \ndcgPPI \label{tab:batch-evaluation-pp-1s}}}
%%%
%%%%%%%%%%%%%%%%%%%%%%%%%%%%%%%%%%%%%%%%%%%%%%%%%%%%%%%%%%%%%%%%%%%%%%%%%%%%%%
  
\subfloat[ProcessNetflow]{ \resizebox{0.7\linewidth}{!}{ \ndcgPNI \label{tab:batch-evaluation-pn-1s}}}\\
%%%
%%%%%%%%%%%%%%%%%%%%%%%%%%%%%%%%%%%%%%%%%%%%%%%%%%%%%%%%%%%%%%%%%%%%%%%%%%%%%%
  
\subfloat[ProcessAll]{ \resizebox{0.7\linewidth}{!}{\ndcgPAI \label{tab:batch-evaluation-pa-1s} }}
%%%
 \label{tab:ndcgScenarioI}
 \end{table} 
%%%%%%%%%%%%%%%%%%%%%%%%%%%%%%%%%%%%%%%%%%%%%%%%%%%%%%%%%%%%%%%%%%%
\begin{table}[tb] 
\caption{Evaluation of batch anomaly scoring in Scenario
  2 (nDCG scores). The higher the score (i.e the closer to 1) the better. The best score per OS (row) is highlighted in bold.}
\centering
\subfloat[ProcessEvent]{ \resizebox{0.7\linewidth}{!}{ \ndcgPEII \label{tab:batch-evaluation-pe-2s}}}
%%%%%%%%%%%%%%%%%%%%%%%%%%%%%%%%%%%%%%%%%%%%%%%%%%%%%%%%%%%%%%%%%%%%%%%%%%%%%%
  
\subfloat[ProcessExec]{ \resizebox{0.7\linewidth}{!}{ \ndcgPXII \label{tab:batch-evaluation-px-2s}}}
\\
%%%
%%%%%%%%%%%%%%%%%%%%%%%%%%%%%%%%%%%%%%%%%%%%%%%%%%%%%%%%%%%%%%%%%%%%%%%%%%%%%%
%  
\subfloat[ProcessParent]{ \resizebox{0.7\linewidth}{!}{ \ndcgPPII \label{tab:batch-evaluation-pp-2s}}}
%%%
%%%%%%%%%%%%%%%%%%%%%%%%%%%%%%%%%%%%%%%%%%%%%%%%%%%%%%%%%%%%%%%%%%%%%%%%%%%%%%
  
\subfloat[ProcessNetflow]{ \resizebox{0.7\linewidth}{!}{ \ndcgPNII \label{tab:batch-evaluation-pn-2s}}}\\
%%%
%%%%%%%%%%%%%%%%%%%%%%%%%%%%%%%%%%%%%%%%%%%%%%%%%%%%%%%%%%%%%%%%%%%%%%%%%%%%%%
  
\subfloat[ProcessAll]{ \resizebox{0.7\linewidth}{!}{ \ndcgPAII \label{tab:batch-evaluation-pa-2s}}}
%%%
 \label{tab:ndcgScenarioII}
 \end{table}

FPOF and OD were not competitive on any dataset, even after trying
several possible support and confidence parameter values and taking
the maximum nDCG score.  The best two methods are AVF and OC3: in scenario 1, AVF produced the best (or tied) results in 8 out of 19 scenarios and OC3 produced the best (or tied)
results, in 12 out of 19 scenarios. In the second attack scenario, AVF produced the best results in 4 out of 20 scenarios and OC3 produced the best results, in 12 out of 20 scenarios. AVF's performance degrades significantly from scenario 1 to scenario 2 (the nDCG range goes from a 0.20-0.84 range in scenario 1 to a 0.17-0.42 range in scenario 2), in particular for the BSD and Android datasets, which might be due to both a large increase in the size of BSD and Android contexts as well as a drop in the percentage of attacks present in the data. AVF performs best on small to medium datasets. In contrast, OC3's performance is more stable between scenarios (the nDCG range goes from 0.21-0.74 in scenario 1 to 0.22-0.84 in scenario 2) and less affected by increase in context size/drop in attack percentage (the performance only really drops for BSD-related contexts and by a smaller margin than in the case of AVF).

 CompreX was not able to complete within a reasonable time; for
wider contexts such as $\ProcessExec$ or $\ProcessParent$, it usually
did not terminate within a few minutes. \citet{comprex}
  mention that CompreX could be run as an anytime algorithm, but the
  available implementation does not support this. In the few cases where CompreX completed in a reasonable time (5 out of 20 scenarios in scenario 1 and 4 out of 20 scenarios in scenario 2), it frequently outperformed both OC3 and AVF and performed best in most cases (3 out of 5 times for scenario 1 and 3 out of 4 times for scenario 2).

In general, nDCG scores were highest for the Android dataset with the first attack scenario (between 0.83-0.84) and
lowest for the Linux dataset, suggesting a rough (but unsurprising)
correlation between the amount of data and difficulty of ranking
attacks effectively.  OC3 and AVF performed considerably better than any other technique on the different
datasets.  Likewise, no single context was
consistently best, and considering all contexts joined together in
$\ProcessAll$ was not always better than considering one of the base
contexts.  

\begin{figure}
  \centering
    \resizebox{0.8\textwidth}{!}{
       \includegraphics[scale=0.3]{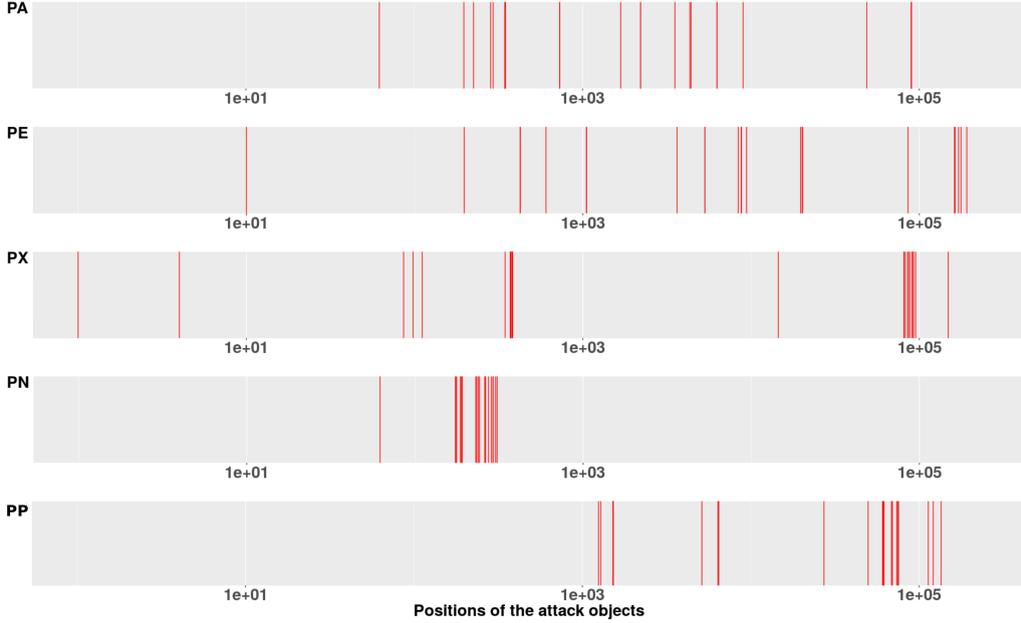}
    }
\caption{Forensic analysis results: Linux (scenario 1)}\label{fig:band-linux}
\end{figure}

To help build intuition regarding how the nDCG scores correspond to
actual rankings, we visualize the results of AVF for Linux (first attack scenario) in
Figure~\ref{fig:band-linux}.  This ``band diagram'' shows the
positions of the attacks in the rankings obtained by AVF for the five
contexts.  The x-axis of the figure is logarithmic scale, so red lines
far to the left represent attacks ranked within the top 10, then top
100, etc.  As this figure illustrates, an nDCG score of 0.43 (obtained
by AVF on the $\ProcessExec$ context in the 1st scenario) corresponds to two attacks found
in the top 10, while scores of under 0.3 tend to correspond to the
highest-ranked attacks occuring at rank 100--1000.

Overall, we can conclude that, AVF and OC3 are \textit{competitive} since they generated the highest nDCG
scores in both scenarios.  

Tables~\ref{tab:time-batch-evaluation-pe-1s}-\ref{tab:time-batch-evaluation-pa-2s}  show the running times for the various algorithms (Tables~\ref{tab:time-batch-evaluation-pe-1s} to \ref{tab:time-batch-evaluation-pa-1s} for Scenario 1 and Tables \ref{tab:time-batch-evaluation-pe-2s} to \ref{tab:time-batch-evaluation-pa-2s} for Scenario 2). Just as with detection performance, the best performing algorithms in terms of running time are OC3 followed by AVF: most scenarios complete under 3 minutes (18 out 20 for both OC3 and AVF for scenario and 19 out of 20 for both OC3 and AVF for scenario 2). Runtime-wise, FPOF, CompreX and OD were significantly
more expensive (in the cases where they complete, they typically run in minutes rather than seconds)
compared to OC3 or AVF. As mentioned previously, Comprex does not complete in a reasonable time in most cases (it only completes in 20\% to 25\% of the cases depending on the scenario). Both OD and FPOF complete in more than 3 minutes in a significant proportion of the cases (7 out of 20 cases for scenario 1 and 14 out 20 cases for scenario 2) so are not competitive in terms of running time as well as detection performance: both algorithms start with frequent itemset/frequent rule mining, which is notoriously computationally  expensive particularly for low support and/or confidence thresholds.

\begin{table}[tb]  
 \caption{Running time results (in seconds) for ProcessEvent context in scenario 1. }

\centering
\label{tab:time-batch-evaluation-pe-1s} 
\begin{tabular}{lccccc}
\toprule
                 & \textbf{FPOF} & \textbf{OD} & \textbf{OC3} &
                                                                \textbf{CompreX} & \textbf{AVF} \\ \midrule
\textbf{Windows} & 47.90         & 57.38       & 0.62   & 60.76            & 0.79         \\ \midrule
\textbf{BSD}     & 3418.85       & 3641.86     & 3.54    & 214.79           & 4.59         \\ \midrule
\textbf{Linux}   & 814.15        & 890.87      & 7.30     & 564.51           & 12.59        \\ \midrule
\textbf{Android} & 0.44          & 0.46        & 0.01       & 13.22      & 0.01        \\ \bottomrule
\end{tabular}

%%%%%%%%%%%%%%%%%%%%%%%%%%%%%%%%%%%%%%%%%%%%%%%%%%%%%%%%%%%%%%%%%%%%%%%%%%%%%%
  
 \caption{Running time results (in seconds) for ProcessExec context in scenario 1. }
\centering
\label{tab:time-batch-evaluation-px-1s} 
\begin{tabular}{lccccc}
\toprule
                 & \textbf{FPOF} & \textbf{OD} & \textbf{OC3} &
                                                                \textbf{CompreX} & \textbf{AVF} \\ \midrule
\textbf{Windows} & 3.65         & 3.62       & 3.72       &DNF         & 12.09         \\ \midrule
\textbf{BSD}     & 28.78      & 26.26     &  2.81&DNF &20.98         \\ \midrule
\textbf{Linux}   & 473.57       & 578.44      & 18.47      &DNF           &84.08        \\ \midrule
\textbf{Android} & 0.46       &0.04       & 0.02     &DNF      & 0.02       \\ \bottomrule
\end{tabular}
%%%
%%%%%%%%%%%%%%%%%%%%%%%%%%%%%%%%%%%%%%%%%%%%%%%%%%%%%%%%%%%%%%%%%%%%%%%%%%%%%%
  
 \caption{Running time results (in seconds) for ProcessParent context in scenario 1. }
\centering
\label{tab:time-batch-evaluation-pp-1s} 
\begin{tabular}{lccccc}
\toprule
                 & \textbf{FPOF} & \textbf{OD} & \textbf{OC3}&
                                                                \textbf{CompreX}  & \textbf{AVF} \\ \midrule
\textbf{Windows} &16.03        & 8.48       & 0.56 &DNF &2.00         \\ \midrule
\textbf{BSD}     & 65.70      & 65.08    & 0.95&DNF& 3.83         \\ \midrule
\textbf{Linux}   &  286.41       & 268.52 & 2.18   &DNF   &14.34        \\ \midrule
\textbf{Android} & NA       &NA       & NA       &NA  & NA       \\ \bottomrule
\end{tabular}
%%%
%%%%%%%%%%%%%%%%%%%%%%%%%%%%%%%%%%%%%%%%%%%%%%%%%%%%%%%%%%%%%%%%%%%%%%%%%%%%%%
  
 \caption{Running time results (in seconds) for ProcessNetflow context in scenario 1. }
\centering
\label{tab:time-batch-evaluation-pn-1s} 
\begin{tabular}{lccccc}
\toprule
                 & \textbf{FPOF} & \textbf{OD} & \textbf{OC3}&
                                                                \textbf{CompreX}  & \textbf{AVF} \\ \midrule
\textbf{Windows} & 0.25         & 0.27       & 1120.22       &DNF         & 156.66        \\ \midrule
\textbf{BSD}     & 36.41      &37.48     & 0.01           &DNF     & 0.02         \\ \midrule
\textbf{Linux}   & 5.80        & 1.99      & 0.12          &DNF     & 0.59        \\ \midrule
\textbf{Android} & 0.06          & 0.01        & 0.009     &DNF        & 0.01        \\ \bottomrule
\end{tabular}
%%%
%%%%%%%%%%%%%%%%%%%%%%%%%%%%%%%%%%%%%%%%%%%%%%%%%%%%%%%%%%%%%%%%%%%%%%%%%%%%%%
  
 \caption{Running time results (in seconds) for ProcessAll context in scenario 1.}
\centering
\label{tab:time-batch-evaluation-pa-1s} 
\begin{tabular}{lccccc}
\toprule
                 & \textbf{FPOF} & \textbf{OD} & \textbf{OC3}&
                                                                \textbf{CompreX}  & \textbf{AVF} \\ \midrule
\textbf{Windows} & DNF         & DNF       & DNF              &DNF  & 2576.82 \\ \midrule
\textbf{BSD}     & 3333.97       & 3632.50     & 57.98         &DNF      & 566.31        \\ \midrule
\textbf{Linux}   & DNF       & DNF      & 181.35            &DNF   & 1951.59        \\ \midrule
\textbf{Android} & 0.191         & 0.24        & 0.03        &DNF     & 0.08        \\ \bottomrule
\end{tabular}
   \end{table} 

%%%
\begin{table}

  \caption{Running time results (in seconds) for ProcessEvent context in scenario 2. }

\centering
\label{tab:time-batch-evaluation-pe-2s} 
\begin{tabular}{lccccc}
\toprule
                 & \textbf{FPOF} & \textbf{OD} & \textbf{OC3} &
                                                                \textbf{CompreX} & \textbf{AVF} \\ \midrule
\textbf{Windows} & DNF        & DNF      & 0.18     & 46.67         & 0.89         \\ \midrule
\textbf{BSD}     & 1840.7       & 2692.35    & 533.96  &  2975.59         & 17.20 \\ \midrule
\textbf{Linux}   & 2768.14       & 6054.92    & 22.77   & 970.79            & 16.16        \\ \midrule
\textbf{Android} & 1551.88       & 1551.88      &0.71 & 45.81  & 0.80      \\ \bottomrule
\end{tabular}

%%%%%%%%%%%%%%%%%%%%%%%%%%%%%%%%%%%%%%%%%%%%%%%%%%%%%%%%%%%%%%%%%%%%%%%%%%%%%%
  
 \caption{Running time results (in seconds) for ProcessExec context in scenario 2. }
\centering
\label{tab:time-batch-evaluation-px-2s} 
\begin{tabular}{lccccc}
\toprule
                 & \textbf{FPOF} & \textbf{OD} & \textbf{OC3}&
                                                                \textbf{CompreX}  & \textbf{AVF} \\ \midrule
\textbf{Windows} & DNF    &DNF      & 4.71         &DNF     & 23.45 \\ \midrule
\textbf{BSD}     & 302.46    &323.59    & 32.72&DNF  &100.07        \\ \midrule
\textbf{Linux}   & 514.98      &513.23      & 42.30     &DNF          &131.05 \\ \midrule
\textbf{Android} & 15.14    &7.29   & 0.51    &DNF   & 1.24     \\ \bottomrule
\end{tabular}
%%%
%%%%%%%%%%%%%%%%%%%%%%%%%%%%%%%%%%%%%%%%%%%%%%%%%%%%%%%%%%%%%%%%%%%%%%%%%%%%%%
  
 \caption{Running time results (in seconds) for ProcessParent context in scenario 2. }
\centering
\label{tab:time-batch-evaluation-pp-2s} 
\begin{tabular}{lccccc}
\toprule
                 & \textbf{FPOF} & \textbf{OD} & \textbf{OC3}&
                                                                \textbf{CompreX}  & \textbf{AVF} \\ \midrule
\textbf{Windows} &DNF      & DNF       & 0.50 &DNF & 2.15        \\ \midrule
\textbf{BSD}     & 326.34    & 280.18  &6.98&DNF &18.46      \\ \midrule
\textbf{Linux}   &  417.15  & 437.27 & 8.92&DNF     &27.63       \\ \midrule
\textbf{Android} & 0.016     &0.015  & 0.02 &DNF      & 0.001  \\ \bottomrule
\end{tabular}
%%%
%%%%%%%%%%%%%%%%%%%%%%%%%%%%%%%%%%%%%%%%%%%%%%%%%%%%%%%%%%%%%%%%%%%%%%%%%%%%%%
  
 \caption{Running time results (in seconds) for ProcessNetflow context in scenario 2. }
\centering
\label{tab:time-batch-evaluation-pn-2s} 
\begin{tabular}{lccccc}
\toprule
                 & \textbf{FPOF} & \textbf{OD} & \textbf{OC3} &
                                                                \textbf{CompreX} & \textbf{AVF} \\ \midrule
\textbf{Windows} &DNF      & DNF      & 0.05   &DNF           & 0.12 \\\midrule
\textbf{BSD}     & 32.76     &36.55     & 1.82  &DNF             & 7.02 \\ \midrule
\textbf{Linux}   & 4.60        & 4.89      & 45.51  &DNF            & 28615.57 \\ \midrule
\textbf{Android} &0.90        & 0.65      & 0.63 &DNF    &0.83       \\ \bottomrule
\end{tabular}
%%%
%%%%%%%%%%%%%%%%%%%%%%%%%%%%%%%%%%%%%%%%%%%%%%%%%%%%%%%%%%%%%%%%%%%%%%%%%%%%%%
  
 \caption{Running time results (in seconds) for ProcessAll context in scenario 2.}
\centering
\label{tab:time-batch-evaluation-pa-2s} 
\begin{tabular}{lccccc}
\toprule
                 & \textbf{FPOF} & \textbf{OD} & \textbf{OC3}&
                                                                \textbf{CompreX}  & \textbf{AVF} \\ \midrule
\textbf{Windows} & DNF         & DNF       & DNF     &DNF           & DNF \\ \midrule
\textbf{BSD}     & DNF       & 3146.47     & DNF  &DNF             & DNF        \\ \midrule
\textbf{Linux}   & DNF       & DNF      & DNF    &DNF         & DNF       \\ \midrule
\textbf{Android} & 1715.82      & 700.81        & 5.92     &DNF      & 16.22 \\ \bottomrule
\end{tabular}
 
\end{table} %

\FloatBarrier

\subsection{Streaming anomaly detection}

In this section, we consider the following empirical questions:
\begin{itemize}
\item Q2: Is the detection performance of streaming AVF competitive
  with batch AVF in terms of nDCG and AUC?
\item Q3: Is the runtime performance of streaming AVF competitive with batch AVF?
\end{itemize}

\subsubsection{Detection performance}
\begin{table*}
%\tiny
 \caption{Summary of the detection performance of batch and streaming AVF on
   $\ProcessAll$ for each dataset, and for block sizes of 1\%, 5\%,
                                10\%, and 25\%.  nDCG and AUC scores (higher is better)  }
%\resizebox{\columnwidth}{!}{
\begin{tabular}[h]{lcccccccc}

\toprule
& \multicolumn{2}{c}{Windows} & \multicolumn{2}{c}{BSD}
                          &\multicolumn{2}{c}{Linux}
                          &\multicolumn{2}{c}{Android}  \\
  \cmidrule{2-9}
&nDCG  &AUC   & nDCG & AUC   &nDCG  &AUC & nDCG&AUC \\

\midrule
Stream 1\%	&0.518	&0.993	&0.524	&0.984	&0.298	&0.927	&0.832	&0.872\\
Stream 5\%	&0.490	&0.984	&0.524	&0.984	&0.298	&0.928	&0.828	&0.857\\
Stream 10\%	&0.522	&0.994	&0.524	&0.984	&0.298	&0.927	&0.826	&0.849\\
Stream 25\%	&0.496	&0.985	&0.525	&0.984	&0.298	&0.928	&0.828    &0.858
\\\midrule
Batch	&0.527	&0.996	&0.524	&0.984	&0.298	&0.927	&0.834	&0.878
\\
  \bottomrule
\end{tabular}
\label{tab:evaluation}
\end{table*}

To evaluate the streaming version of AVF, we generated 10
randomly-shuffled versions of each dataset from Scenario 1 and ran the streaming
algorithm on each dataset.  We consider different
randomly-shuffled datasets in order to avoid any dependence on a
particular order of processing the data; it could be that analyzing
the data ordered by time could produce better (or worse) results.  In
practice, it is not guaranteed that we will see all processes in
temporal order, because records for some long-lived processes may not
become available until the process terminates. 
We divided the datasets into block sizes
of various granularities (1\%, 5\%, 10\%, 25\% of the data) to investigate the
effect of granularity on effectiveness and performance. 
For each dataset and block size, we computed the median ranking of each attack over
the 10 shuffled runs.  These median rankings are taken to be
representative.

We present nDCG and AUC results for the
$\ProcessAll$ context only; these results are representative of the
base contexts.  Table~\ref{tab:evaluation} summarizes the nDCG and
AUC metrics for the streaming algorithm (with four different
block sizes) and for the batch algorithm (at the bottom).  These
results show that the nDCG scores for all four  datasets
are fairly stable, with only the Windows dataset displaying
degradation of nDCG score of more than 0.01.  Likewise, the AUC scores
of most streaming variants were close to those of the
batch algorithm, with only the Windows and Android AUC scores
changing by more than 0.01.  Overall these results suggest that
small block sizes do not significantly degrade the usefulness of the
results of AVF scoring.

\begin{figure*}%
   \subfloat[Windows]{\label{fig:ntp-pe-win}
      \includegraphics[width=0.5\textwidth]{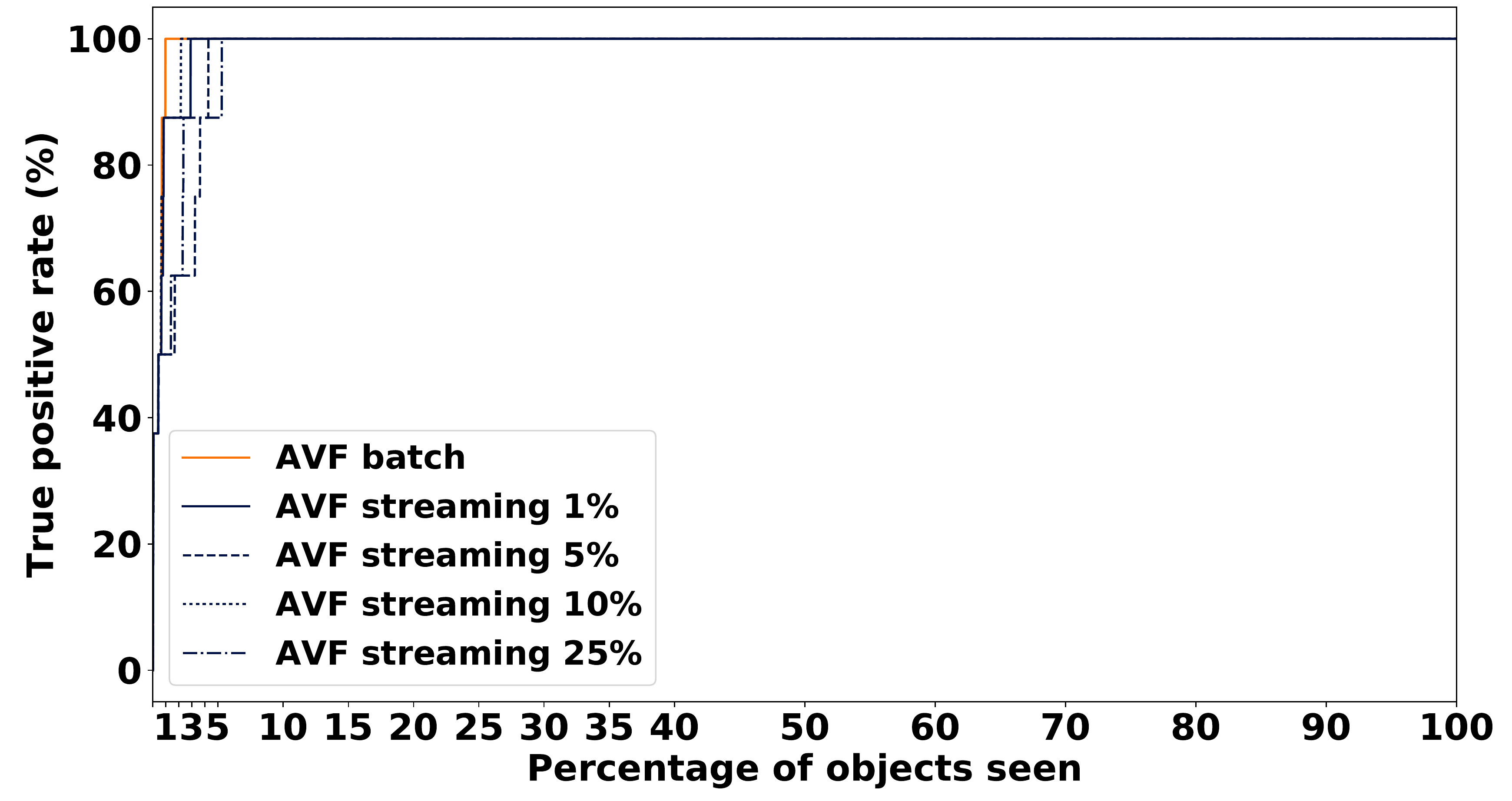}}
   \subfloat[BSD ]{\label{fig:ntp-pe-bsd}
      \includegraphics[width=0.5\textwidth]{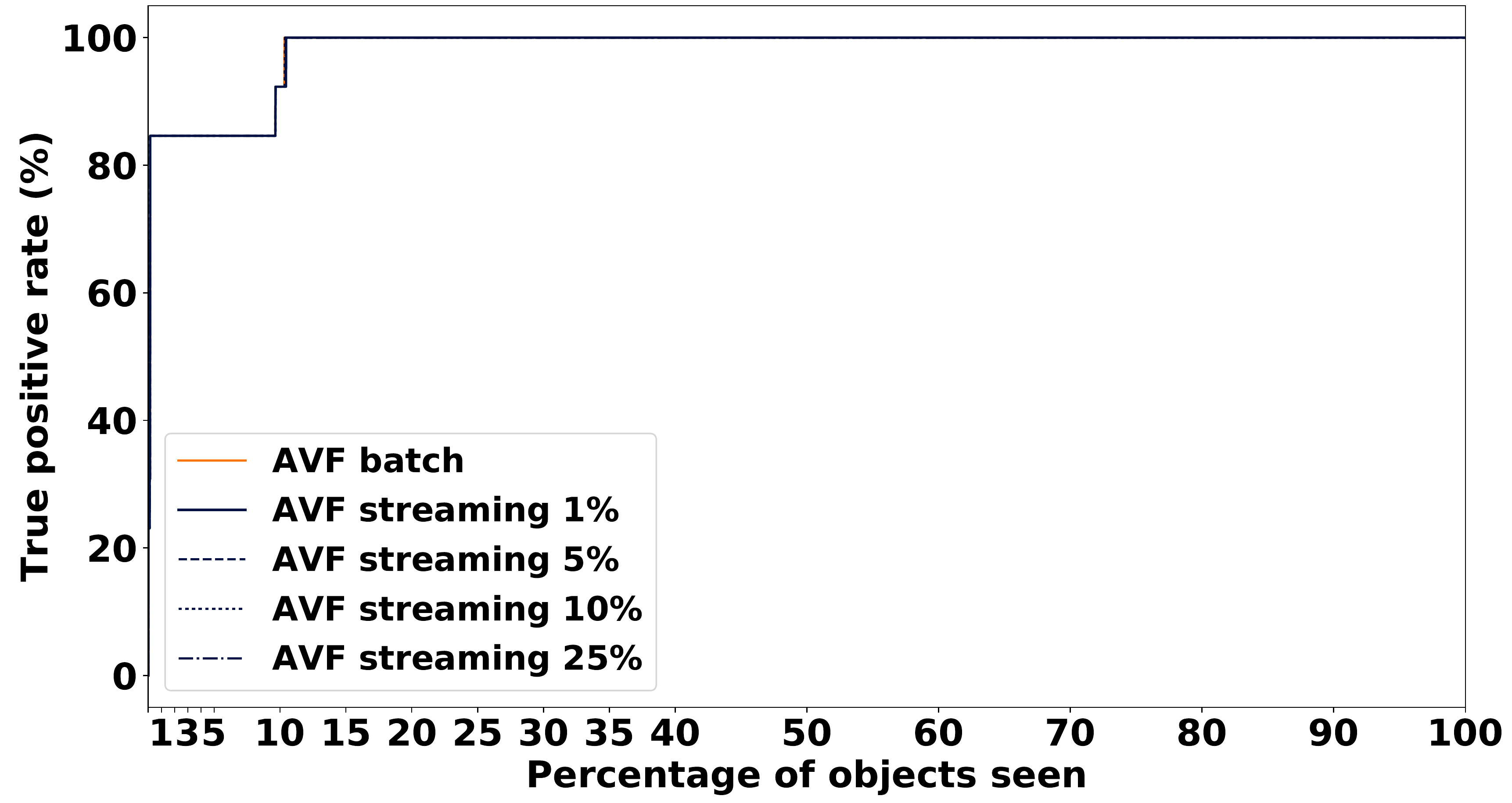}}

   \subfloat[Linux]{\label{fig:ntp-pe-linux}
      \includegraphics[width=0.5\textwidth]{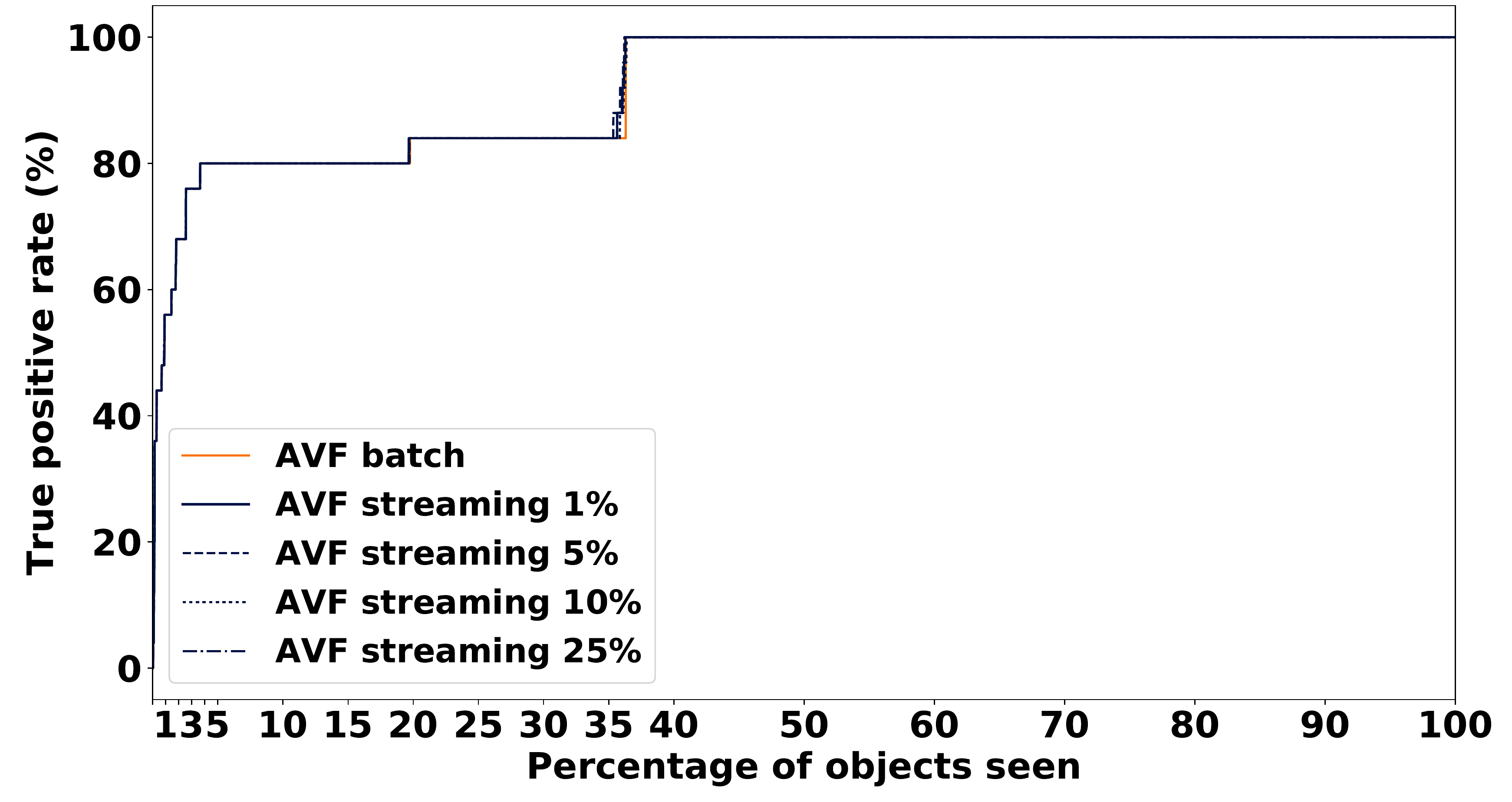}}
   \subfloat[Android]{\label{fig:ntp-pe-android}
      \includegraphics[width=0.5\textwidth]{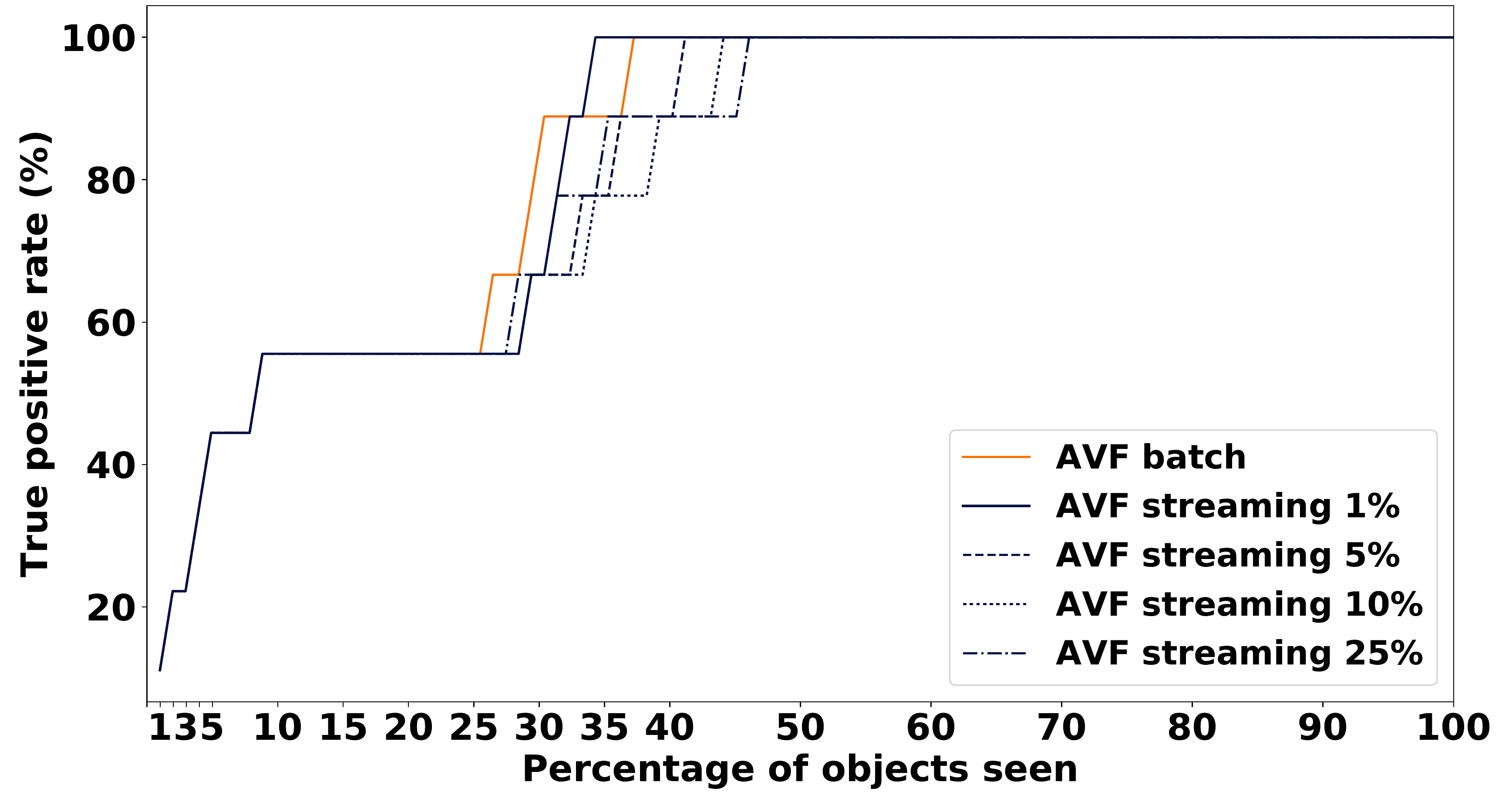}}
    \caption{Percentage of processes seen versus percentage of attacks
      detected  for $\ProcessAll$}%
    \label{fig:ntp-pa}%
\end{figure*}

Figure~\ref{fig:ntp-pa} plots the ratio of true positives found
vs. ranking position, for the four different $\ProcessAll$ datasets.
The red lines are the performance of the batch AVF algorithm while the
blue lines are the streaming versions.  (For the BSD dataset, the
differences are not visible.)  We can also gain a stronger intuition
regarding the usefulness of the results from these figures: for
example, for the Linux $\ProcessAll$ context we can see that the nDCG
score of 0.298 corresponds to finding about half of the attacks in the
first 1\% of the rankings, while others are not found until
40\%. Figure~\ref{fig:ntp-pa} also shows that, for most datasets (except Android), at least 80\% of true positives (i.e attacks) are found in the top 5\% of the data. 

% Figures~\ref{fig:ntpWin}--\ref{fig:ntpOther} (in particular for contexts $\ProcessNetflow$ with Linux, $\ProcessExec$ with Windows and $\ProcessEvent$ with Android) 
% show that AVF is more stable than AVC when block size varies. They also show that AVF performs better than AVC for smaller datasets and that AVC's 
% performance degrades significantly the smaller the block size gets. This is consistent with the fact that AVC is a compression-based method and would perform better with more information and regular patterns.
% AVC still performs better than AVF on larger datasets (though here as well the performance degrades with the shrinking of the block size) as shown for example, for the Linux $\ProcessAll$ context where AVC manages
% to detect all attack processes with only 3\% of the available data.

\subsubsection{Analysis time}

\begin{figure*}%
   \subfloat[Windows]{\label{fig:5dbatch}
      \includegraphics[width=0.5\textwidth]{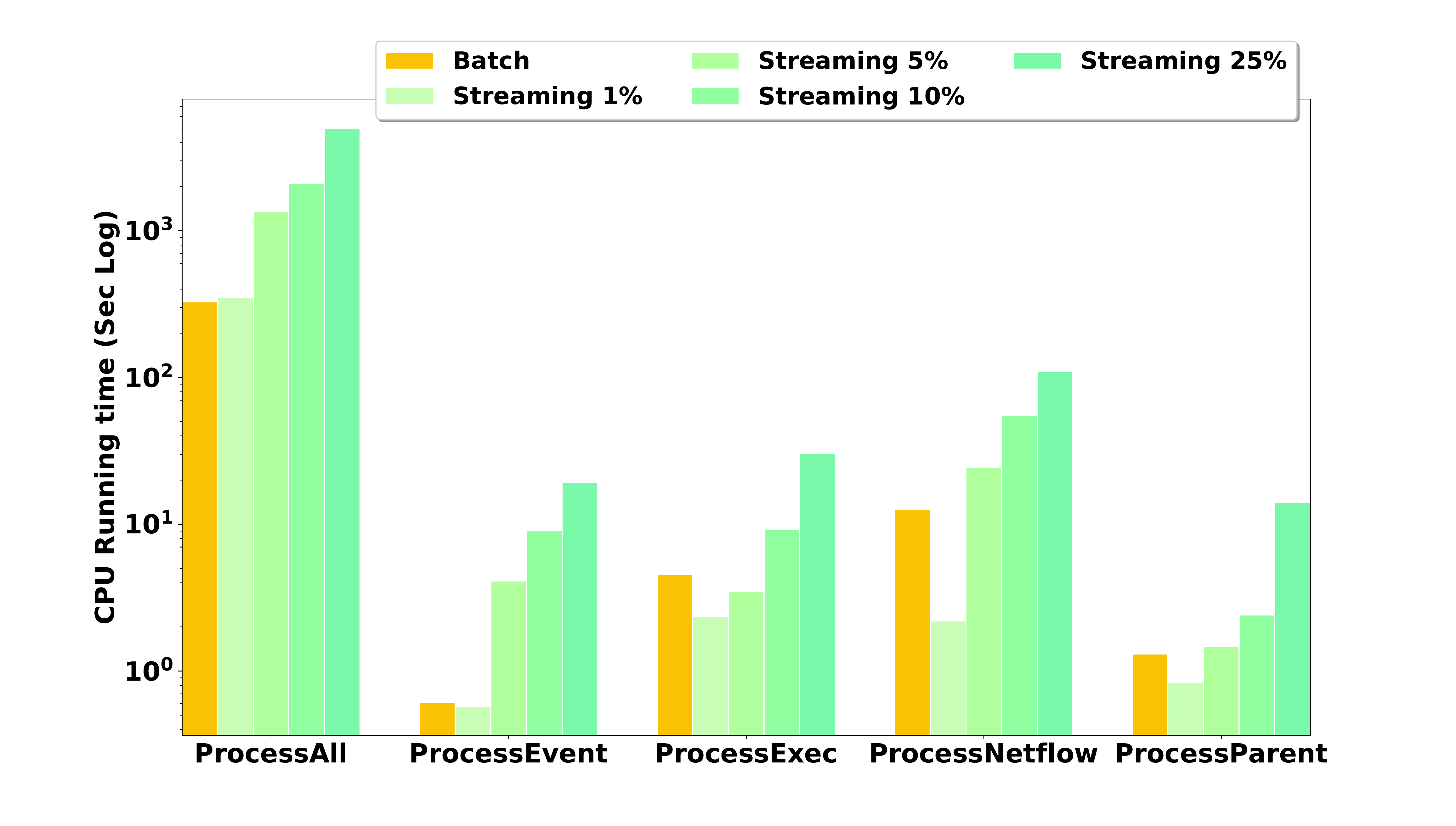}}\hfill
      \subfloat[BSD]{\label{fig:cadetsbatch}
      \includegraphics[width=0.5\textwidth]{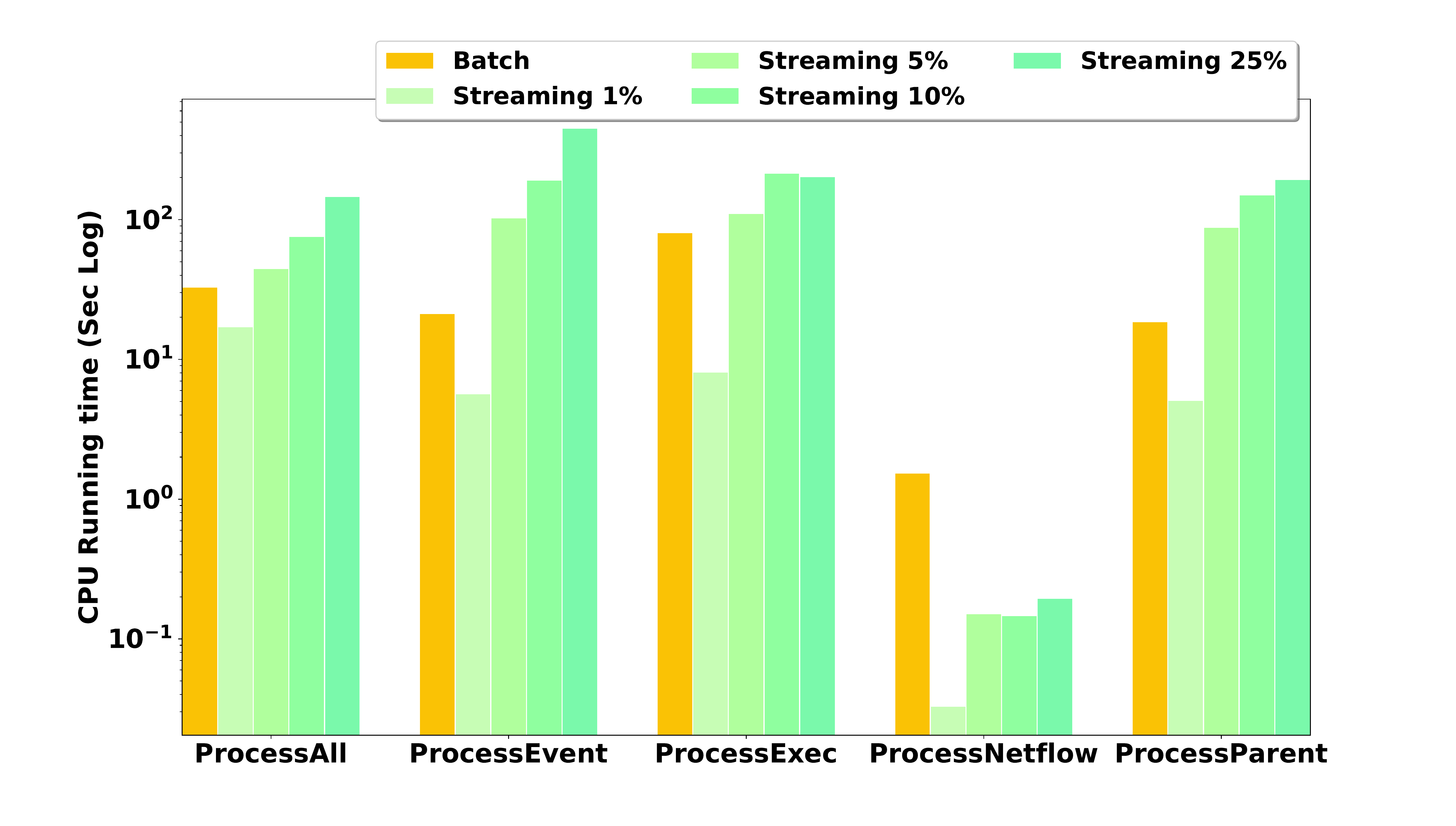}}
\\
   \subfloat[Linux]{\label{fig:tracebatch}
      \includegraphics[width=0.5\textwidth]{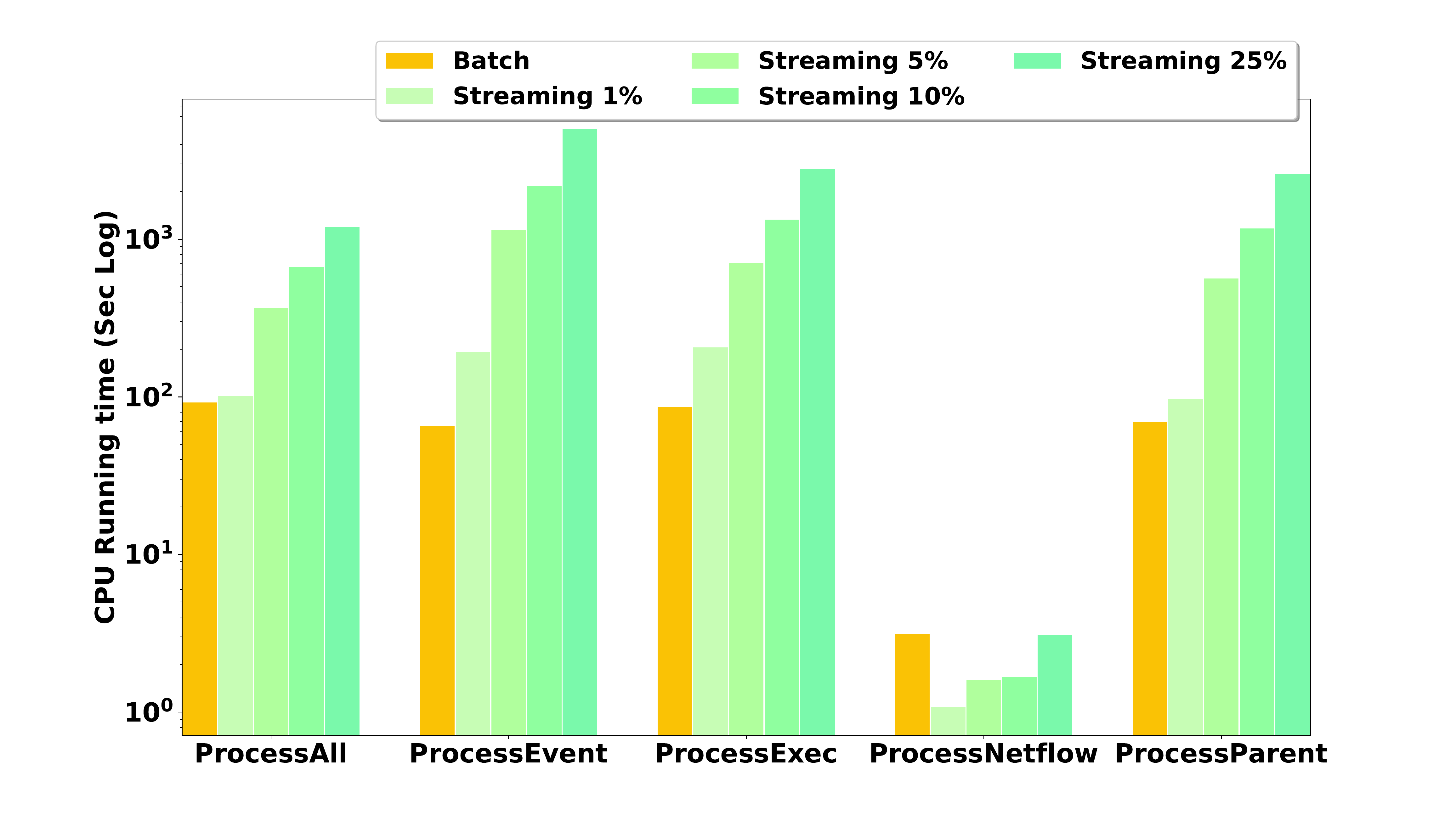}}\hfill
   \subfloat[Android]{\label{fig:clearscopebatch}
      \includegraphics[width=0.5\textwidth]{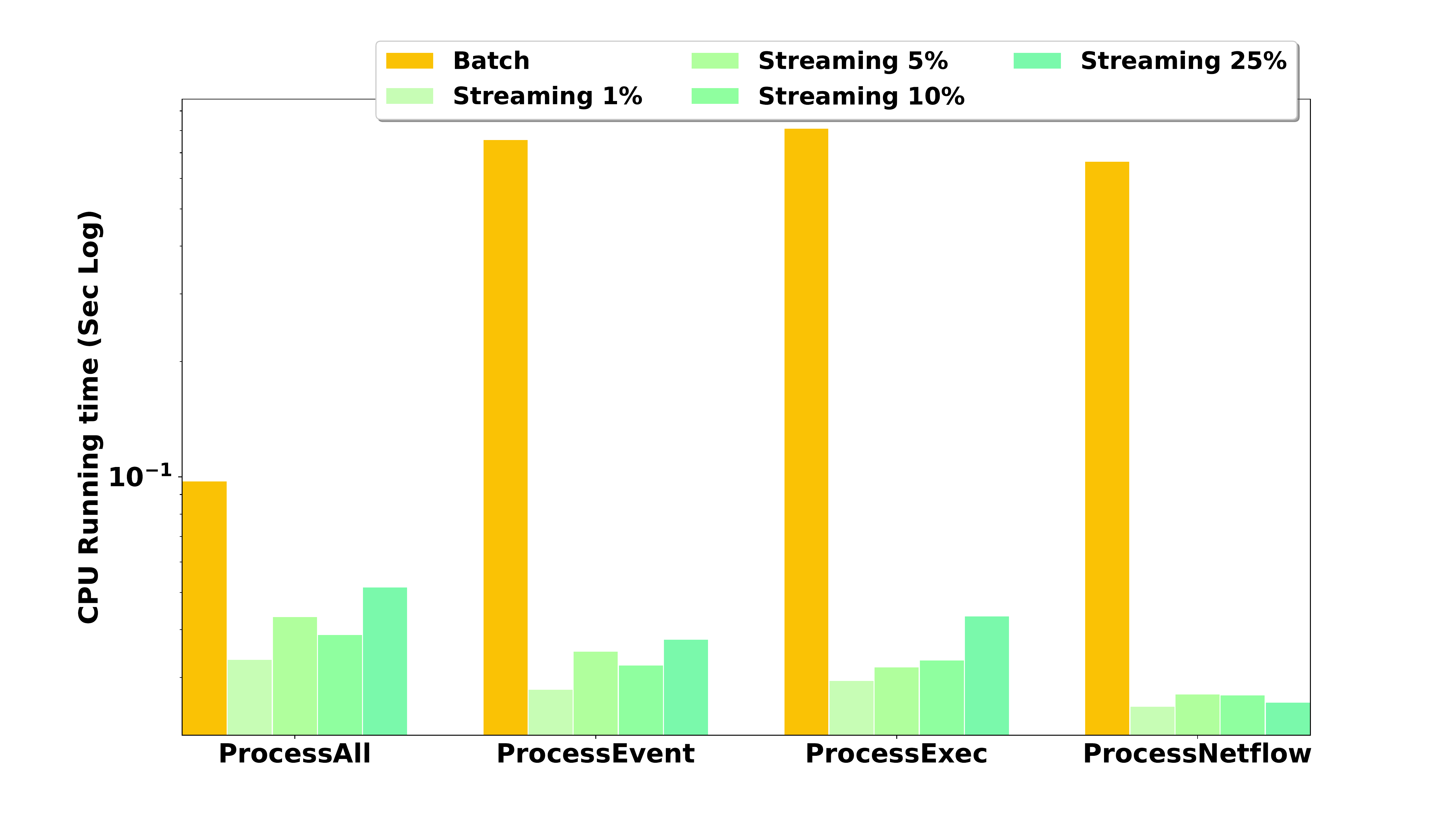}}
%     \begin{subfigure}{0.6\textwidth}
%     \includegraphics[width=\textwidth]{fig/5dbatch.pdf}
%     \caption{Windows forensic analysis time}\label{fig:5dbatch}
%     \end{subfigure}
%     \begin{subfigure}{0.6\textwidth}
%     \includegraphics[width=\textwidth]{fig/cadetsbatch.pdf}
%     \caption{BSD forensic analysis time}\label{fig:cadetsbatch}
%     \end{subfigure}\\
%      \begin{subfigure}{0.6\textwidth}
%     \includegraphics[width=\textwidth]{fig/tracebatch.pdf}
%     \caption{Linux forensic analysis time}\label{fig:tracebatch}
%     \end{subfigure}
%      \begin{subfigure}{0.6\textwidth}
%     \includegraphics[width=\textwidth]{fig/clearscopebatch.pdf}
%     \caption{Android forensic analysis time}\label{fig:clearscopebatch}
%     \end{subfigure}
    \caption{Analysis time (batch AVF vs. streaming AVF)}%
    \label{fig:forensic-time}%
\end{figure*}

Figure~\ref{fig:forensic-time} summarizes the time taken per run for both batch and streaming versions of AVF (the streaming times were obtained by taking the median of the times over the ten runs on shuffled inputs). Note that the y-axis is logarithmic scale.
The running time is in general proportional to the amount of data in each context (number of rows $\times$ number of columns).
In particular, the time needed for $\ProcessAll$ is often considerably longer than the times needed for the other contexts. The reason is that some contexts (such as $\ProcessEvent$) have many rows and few columns,
while others (such as $\ProcessNetflow$) have many columns and few rows. Combining them into $\ProcessAll$ yields a very sparse context with many zeros. We plan to investigate whether using a more
succinct storage format for the contexts, or combining the scores of the subcontexts, might lead to better performance. The streaming execution times also increase, as expected, with the increase of streaming block size.

\section{Related work}
\label{sec:relatedwork}
%section about FCA/ARM and potentially close applications to our own (anomaly detection with these two?)
%malware detection papers
%graph anomaly detection section

Prior work on APTs is mostly concerned with describing/modeling the
characteristics of an APT and its attack
model~\cite{sood2013targeted,virvilis2013trusted,chen2014study},
sometimes using case studies~\cite{Karchefsky2017}. % Yang et
% al.~\cite{yang2017maximizing} seeks to improve the understanding of
% APT characteristics through modeling the APT effectiveness
% maximization problem as a constrained optimization problem.  
A few recent studies address the APT detection problem by constructing
models of normal behavior against which incoming data is compared and
flagged as anomalous if it deviates from the learned models. \citet{friedberg2015combating} explain the shortcomings of
current security solutions with regards to APT detection, in
particular contending that preventive security mechanisms and
signature-based methods are not enough to tackle the challenge of
APTs, and propose an anomaly detection-based framework to detect APTs
by learning a model of normal system behavior from host-based security
logs and detecting deviations.  \citet{siddiqui2016detecting} use the fractal dimension as a
feature to classify TCP/IP session data patterns into anomalous (and
part of an APT) or normal patterns.  \citet{moya15expert}
construct decision tree-based models of normal
network activity based on features extracted from firewall logs, then
use the learned models to classify incoming network
traffic.  % A very recent study~\cite{brogi2017hidden} models APT phases
% with a Hidden Markov Model (HMM) afterwards used to characterize
% potential attack campaigns as likely or unlikely to be part of an APT.
% Other studies try to model advanced persistent threats based on game
% theory so as to suggest optimal mitigation
% strategies~\cite{hu2015dynamic,pawlick2015flip,rass2017defending}.
Some work has also been done on the detection of specific patterns
that might be part of an APT attack e.g. detection of data
leakage/data exfiltration~\cite{jewell2011host,awad2016data} or
detection of command and control (C\&C)
domains~\cite{niu2017identifying}. Another recent paper~\cite{lamprakis2017} reconstructs a Web requests dependencies graph from Web requests logs using domain knowledge and proposes an unsupervised approach relying on the reconstructed graph to identify APT C\&C channels.
In contrast, in this paper, we seek to evaluate APT detection approaches developped on host-based data (unlike ~\cite{lamprakis2017,niu2017identifying,siddiqui2016detecting,moya15expert} that rely on datasets recording various aspects of network activity) that use as little domain knowledge as possible (the goal being to check the detection performance on datasets constructed to minimize the amount of pre-processing and fine-tuning) and try to detect traces of APT activity without targeting a specific type of APT pattern (unlike~\cite{jewell2011host,awad2016data}).

There is a considerable literature on intrusion and malware
detection, which is mainly split in two approaches: misuse detection
(e.g.~\cite{kumar1994pattern}) and anomaly detection
(e.g. \cite{ji2016multi}).  The principle of misuse detection is to
search for events (i.e. known attacks) that match predefined signatures
and patterns. Methods relying on misuse detection can only detect
attacks whose signature and patterns are known, which would be
unsuitable for APT detection. By contrast, anomaly detection assumes
abnormal behaviours can come in varied, potentially unknown, shapes
and focuses on detecting activity that deviates from normal activity
i.e. activity usually recorded on a particular host or network.  % Sabahi
% and Movaghar~\cite{sabahi2008intrusion}, Liao et
% al.~\cite{liao2013intrusion} and 
% Zuech et
% al.~\cite{zuech2015intrusion} provide a recent survey of intrusion detection
% techniques while Agrawal and Agrawal~\cite{agrawal2015survey} surveys
% data mining techniques for intrusion detection.  
% Ahmed et
% al.~\cite{ahmed2016survey} is another survey on the main network
% anomaly detection techniques, which it divides into four categories:
% classification-based anomaly detection, clustering-based anomaly
% detection, information theory-based anomaly detection and statistical
% anomaly detection.

% Among the many papers describing applications of Formal Concept
% Analysis (FCA) and frequent itemsets mining (see~\cite{Poelmans2010}
% for a detailed survey of the FCA applications), few describe
% security-related applications. Of these, most make use of FCA and
% frequent itemset mining to detect software code defects and
% anomalies~\cite{li2005pr,wasylkowski2007detecting}. Fredrikson et
% al.~\cite{fredrikson2010synthesizing} is, to our knowledge, the only
% study that directly makes use of FCA for malware detection: in this
% case, FCA is used to refine learned models/discriminative
% specifications of malware/normal system behaviour.

% \textbf{This is in a supervised setting, unlike us.}

% \textbf{TODO: AD for itemsets and categorical data: FPOutlier, Krimp, CompreX}

There are several comprehensive surveys of anomaly detection and
outlier detection that consider categorical data, continuous data, and
structured data (e.g. graphs)~\cite{anomaly,graph-anomaly}.  Of these
approaches, graph anomaly detection appears the most relevant for our
problem, but most of this work has considered special cases of graphs
(e.g. undirected or unlabeled), whereas provenance graph data has rich
structure (labeled nodes, labeled edges, multiple properties on nodes
and edges).  Anomaly detection approaches for provenance graphs
reported so far rely on training on benign traces~\cite{streamspot},
require user-provided annotations~\cite{sleuth}, or assume that the
background activity is highly regular~\cite{winnower}.  Another recent
contribution by \citet{siddiqui18kdd} shows that human-in-the-loop feedback can
be used in a semi-supervised way to improve detection results over
baseline unsupervised detectors over numerical data.
\citet{tapp2019} investigated \emph{aggregation} of anomaly
scores/ranks from different contexts, and found that using AVF and OC3
as base detectors, simple score or rank aggregation techniques provide
improved detection performance.  
% (Interestingly, the
% Winnower system~\cite{winnower} based on graph grammar induction can
% arguably be viewed as an instance of a MDL approach already, since it
% builds a compact summary of the regular structure of the data by
% learning graph grammar rules, and highlights as anomalies those parts
% that are not captured by the rules.)

On the other hand, there are a number of generic approaches to anomaly
detection for discrete (categorical)
data~\cite{fpoutlier,outlierdegree,avf,ndi-od,krimp-ad,upc,comprex,upc}.
Most of these approaches first mine the data for frequent itemsets or
association rules, and all then perform anomaly scoring in a second
pass over the data.  A one-pass, streaming variant of AVF was
presented by ~\citet{onepassavf}. Some approaches,
notably OC$^3$~\cite{krimp-ad} and CompreX~\cite{comprex}, are based
on the Minimum Description Length (MDL) principle~\cite{mdl}.  Both
perform a preprocessing stage to find a compressed representation of
the dataset, then consider the resulting compressed size of each
record as its score.  Since OC$^3$ was often the most effective batch
algorithm, we think it would be interesting to develop a streaming
approach based on MDL, either by adapting the underlying Krimp
compression algorithm~\cite{krimp} to support streaming anomaly
detection, or by building on streaming compression techniques such as
adaptive arithmetic coding~\cite{witten87cacm}.  The UPC algorithm of
\citet{upc} is also based on pattern mining and MDL, and is inherently
a two-pass approach, but seeks a different kind of
anomalies than AVF, OC3, and CompreX, consisting of unexpectedly rare
combinations of frequent itemsets.
% Our AVC approach is also based on MDL but uses a
% simpler model that is easy to maintain incrementally; on the other
% hand, our approach may miss opportunities for compression when
% attributes are not independent.  Further work is needed to compare our
% results with what would be possible using these more sophisticated
% offline analyses, and devise a hybrid streaming approach that could take
% advantage of correlated attributes.

There are also some anomaly detection techniques for mixed categorical
and numerical data~\cite{smartsifter,odmad} that could be applied to
pure categorical data.  The ODMAD algorithm~\cite{odmad}, like most
categorical techniques, performs an initial off-line pattern mining
stage.  To the best of our knowledge SmartSifter~\cite{smartsifter} is
the only previous unsupervised online algorithm applicable to
categorical data.  SmartSifter incrementally maintains a histogram
density model of the categorical data and, for each combination of
attributes, a continuous distribution (such as a multivariate Gaussian
mixture model) for the numerical attributes.  % As in our approach, the
% score for a record $x$ is $-\log p(x)$ where $p(x)$ is the predicted
% probability of $x$ given past data, so theirs can also be viewed as an
% MDL-based approach.  
SmartSifter's running time is $O(2^m d^2 k)$
where $m$ is the number of categorical attributes, $d$ the number of
numerical attributes (i.e. dimension) and $k$ the number of components
of the mixture model. Their experiments considered datasets with
$m \leq 1$ and $d \leq 7$, and it is unclear whether this approach can
scale to large numbers ($m > 100$) of categorical attributes.  The
techniques based on itemset mining are exponential in the number of attributes
in the worst-case, but have acceptable performance in practice, while
the AVF approaches require only $O(m)$ time to process each input record.

Relatively few publications making use of the DARPA Transparent
Computing datasets have appeared; much of the data has not been made
publicly available, and ground truth annotations are often not
available in machine-readable form.  In some cases, systems have been
evaluated using these datasets but the raw data, or derived products,
have not been made available, making it difficult to reproduce their
results.  
Both \citet{siddiqui18kdd} and
\citet{tapp2019} used datasets derived from Transparent
Computing but the data were not made publicly available.  We believe that this article is the first to evaluate
anomaly detection algorithms on publicly available datasets derived from the Transparent
Computing project.

% \subsection{APT detection and malware detection}
% \subsection{Anomaly detection in graphs}
%\subsection{}

\section{Conclusion}

\label{sec:conclusion}

Detecting APT-style attacks in real-world settings is extremely
difficult in general.  In this paper, we investigate the feasibility
of finding processes that may be part of such attacks by analyzing
their behavior.  We considered five different batch algorithms, one of
which can also be adapted easily to a streaming setting.  Our
experiments showed that both batch and online approaches are effective
in finding attacks and can analyze several days' worth of activity
(tens or hundreds of thousands of process summaries, sometimes with
over ten thousand attributes) in a few minutes, a negligible cost
compared to the time and effort needed to record and store this data.
Moreover, our results are validated on provenance traces gathered from
four different operating systems, subject to several different kinds
of attacks; many of the attacks were typically ranked among the top
0.1-1\%.

We believe that this work represents a significant contribution, in
that it can provide a low-cost, yet effective line of defense in a
larger provenance-based monitoring system, and establishes a baseline
for comparison of more sophisticated (and time-consuming) techniques.
Nevertheless, there are a number of areas for improvement.  First,
interpreting and analyzing the processes flagged for investigation is
still mostly a manual process, motivating further support for
identifying connections between the most anomalous processes.  Second,
it is also important to consider the (common) case when there is no
attack.  Since attacks are rare and, in a given trace, there are
typically hundreds or thousands of anomalous processes that are not
part of the attack, more work is needed to identify suitable
thresholds to limit effort in this case.  Finally, our approach
assumes that the attacker is not aware of or able to manipulate the
detection system; sophisticated attackers will naturally seek to
either evade observation entirely or modify their behavior so as to
minimize anomaly scores.  Further research is needed on how to
make anomaly detection robust even if attackers know how their
activity is being monitored.

\section*{Acknowledgements}
This material is based upon work partially supported by the Defense Advanced Research Projects Agency (DARPA) under contract FA8650-15-C-7557.
Mookherjee was partially supported by a grant from LogicBlox, Inc.
%If you'd like to thank anyone, place your comments here
%and remove the percent signs.

% BibTeX users please use one of
\bibliographystyle{plainnat}      % basic style, author-year citations
\bibliography{biblio}   % name your BibTeX data base
% \appendix
% \section{Additional experimental results}

\printglossaries

\end{document}